\documentclass[electronic]{vgtc}               

\graphicspath{{figures/}{pictures/}{images/}{./}} 

\usepackage{times}                     

\usepackage{tabu}                      
\usepackage{booktabs}                  
\usepackage{lipsum}                    
\usepackage{mwe}                       

\usepackage{mathptmx}                  
\usepackage{makecell}
\usepackage{graphicx}

\usepackage{array}
\usepackage{multirow}
\usepackage[table]{xcolor}
\usepackage[utf8]{inputenc}
\usepackage{tabularx}
\usepackage{pifont}

\onlineid{xxxx}

\vgtccategory{Research}

\vgtcinsertpkg

\title{HandPad: A Bimanual Hand Interface for Fluid Window Interactions in VR}

\author{
\href{https://wy-blacksheep.github.io/}{Wen Ying}$^{1}$,
\href{https://adildsw.com/}{Adil Rahman}$^{1}$,
\href{https://erzhenh.com/}{Erzhen Hu}$^{1}$,
and
\href{https://seongkookheo.com/}{Seongkook Heo}$^{2}$
}

\newcommand{\projectpagept}{10}
\newcommand{\projectpagebaselineskip}{12}
\newcommand{\projectpagefont}{%
  \fontsize{\projectpagept pt}{\projectpagebaselineskip pt}\selectfont
}
\newcommand{\projectpageafterskip}{3pt}

\affiliation{\scriptsize
$^{1}$Department of Computer Science, University of Virginia\\
$^{2}$Department of Computer Science and Engineering, Ulsan National Institute of Science and Technology\\[2ex]
{\projectpagefont Project page: \href{https://handpad.github.io/}{\textcolor[HTML]{455fd4}{https://handpad.github.io/}}}
\vspace{\projectpageafterskip}
}

\teaser{
  \centering
  \includegraphics[width=\linewidth]
  {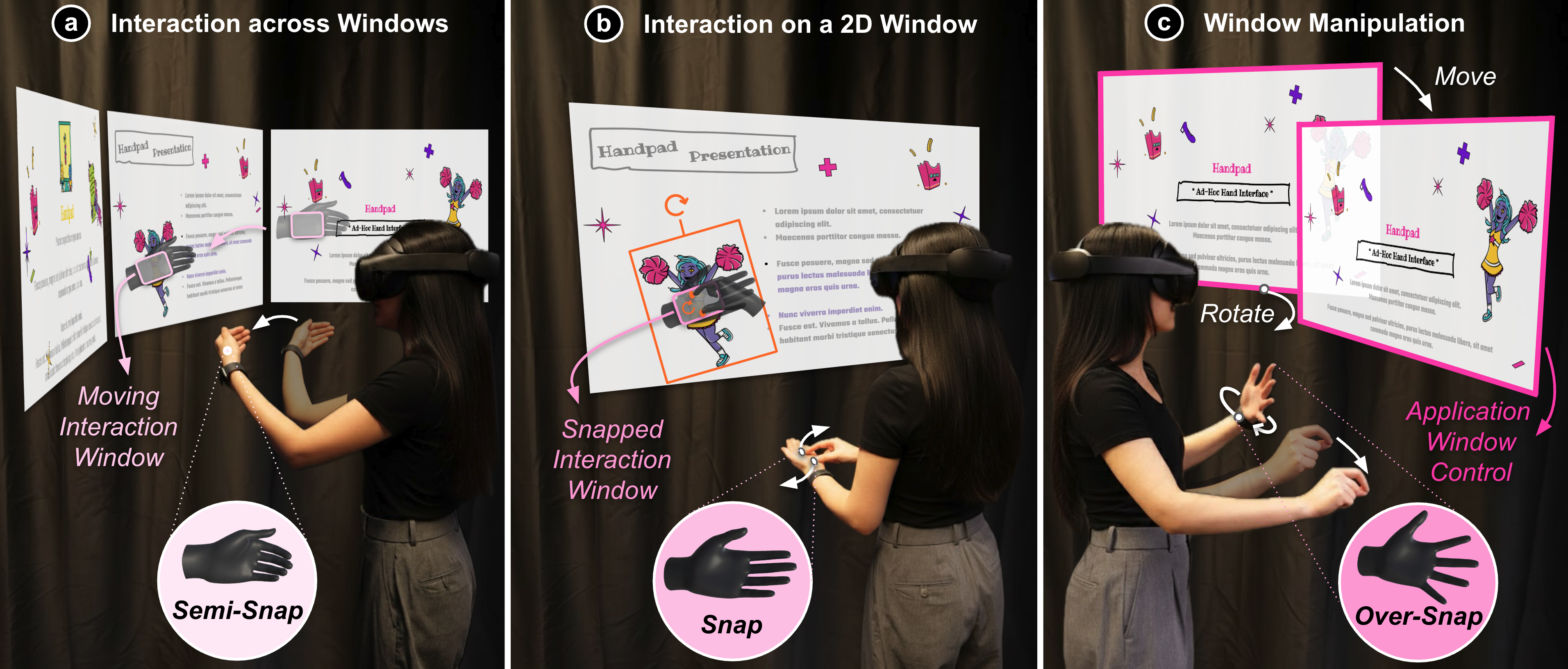}
  \vspace{-5mm}
  \caption{HandPad utilizes the non-dominant hand (NDH) to establish spatial frames and interaction contexts, enabling the dominant hand (DH) to perform fine-grained manipulations. (a) Semi-Snap: A slightly bent NDH gesture initiates an interaction window, serving as a spatial reference for navigating between applications. (b) Snap: A Flat NDH fixes the interaction window, enabling the DH to perform direct and precise interactions on the palm. (c) Over-Snap: A widely open NDH transitions the system to coarse-grained application management. Users can adjust window orientation via NDH rotation and relocate the window using DH pinch gestures. Both physical hands are remapped to the virtual window to mimic direct manipulation while improving ergonomics.}
  \label{fig:teaser}
}

\abstract{
Virtual Reality (VR) offers potential for productivity work by creating expansive displays anywhere, yet current systems often rely on external input devices that limit the on-the-go use of mobile VR. We introduce HandPad, a suite of bare-hand interaction techniques that leverage the benefits of asymmetric bimanual coordination and self-haptic support. HandPad assigns the non-dominant hand (NDH) to establish spatial frames and interaction contexts, while the dominant hand (DH) performs fine-grained manipulation.
Users can use NDH gestures as an input modifier to change the mode and target of DH interactions, including multi-window navigation, in-window content interaction, and window management. The palm surface of the NDH also serves as a physical touch surface, providing passive haptic feedback for effective DH touch interaction. Both hands and their interactions are spatially remapped to the window surface, enabling comfortable and direct interaction with virtual content.
An exploratory study showed that HandPad enables efficient and ergonomic interaction, demonstrating its potential as a device-free approach for knowledge work in VR.
} 

\keywords{Gestures, bimanual interaction, input remapping, virtual reality.}

\nocopyrightspace

\makeatletter
\usepackage{etoolbox}
\patchcmd{\@maketitle}{%
           \else%
	    \vspace{1\baselineskip}\fi%
           \large\sffamily\vgtc@sectionfont}{%
           \else%
	    \vspace{1\baselineskip}\fi%
           \large\sffamily\vgtc@sectionfont}{}{}
\makeatother

\makeatletter
\patchcmd{\@maketitle}{%
               \par\vspace{1\baselineskip}%
               \vgtc@affiliation\par}{%
               \par\vspace{0.5\baselineskip}%
               \vgtc@affiliation\par}{}{}
\makeatother

\begin{document}


\maketitle



\section{INTRODUCTION}
Virtual Reality (VR) opens new possibilities for productivity work, offering customized environments that transcend physical constraints such as limited display and ambient noise~\cite{10.1145/3313831.3376724, 8617763}. However, standard VR inputs, such as gaze, controllers or in-air gestures, often lack the precision and haptic feedback required for information work on 2D windows~\cite{10.1145/3025453.3025474, simeone2023everyday}. While researchers have integrated conventional input tools like mice or tablets into VR~\cite{8943608, 9284682}, these hardware-dependent solutions often reduce portability in mobile VR settings and are difficult to adapt to versatile spatial displays~\cite{ofek2020practical}. Overall, efficiently interacting with large, multi-window environments without dedicated hardware remains a critical challenge for the on-the-go use of mobile VR.

Advances in hand tracking have integrated bare-hand interaction into commercial headsets like the Meta Quest 3 and Apple Vision Pro; however, these systems typically assign both hands identical functions, overlooking the dexterity and specialized manual roles of the two hands~\cite{jones2006human}. In practice, interacting with virtual windows is a joint activity that requires both stability and precision, a coordination that can benefit from asymmetric bimanual interaction~\cite{guiard1987asymmetric}. In this model, the non-dominant hand (NDH) establishes the spatial frame and interaction context, while the dominant hand (DH) performs fine-grained actions and detailed content manipulation relative to that established reference~\cite{guiard1987asymmetric, hinckley1997attention}. However, standard touch interfaces often struggle to support such asymmetric bimanual input because they lack the interaction context to distinguish between the hands~\cite{webb2016wearables}. VR provides a more suitable setting for this asymmetric bimanual interaction model, as it can continuously track both hands and interpret their spatial relationship during input. We therefore hypothesize that assigning differentiated labor to each hand and enabling them to coordinate~\cite{hinckley1994passive, two1994kabbash, hinckley1997cooperative, role1999balakrishnan} can facilitate more intuitive workflows and improve the productivity of 2D window interactions for productivity work in VR.

To truly leverage the benefits of bimanual asymmetry~\cite{guiard1987asymmetric}, we propose HandPad, a collection of bare-hand bimanual interaction techniques for direct interaction with virtual content across multiple application windows. With HandPad, the user's NDH is leveraged as (1) a \textbf{spatial frame} to establish the interaction space (i.e., interaction window in Figure~\ref{fig:teaser}) and providing a reference to anchor the DH's motions; (2) a \textbf{tangible surface} stabilizing the frame to support precise DH touch interactions on the NDH via a machine-learning touch detection method (Figure~\ref{fig:teaser}b); (3) a kinesthetic \textbf{mode-switching modality} using gesture to phrase interactions and create context-specific meanings for different inputs such as manipulating contents or windows. This division of manual labor follows the principle of Left-Hand Precedence~\cite{guiard1987asymmetric, webb2016wearables}, where the NDH initiates the interaction context while the DH performs fine-grained manipulations within that frame. With HandPad, the NDH supports three distinct modes designed to match the different demands of multi-window interaction in VR. Semi-Snap for navigating across windows, Snap for establishing a touch interaction area, and Over-Snap for managing application windows. HandPad also addresses the ``gorilla arm'' issue commonly found in bare-hand interfaces~\cite{hansberger2017dispelling} by allowing users to manipulate at a lower, more comfortable physical position and remapping their hand and inputs to higher virtual windows~\cite{feuchtner2018ownershift, medeiros2023benefits, ying2026redirected}. In addition, HandPad techniques are fully compatible with current VR gaze and mid-air gestures. Users can employ gaze to coarsely position the interaction window and use mid-air gestures to modulate its size or location (Figure~\ref{fig:teaser}c). We conducted a study to examine the benefits of HandPad for virtual window interaction in VR. Our results show that HandPad particularly benefits tasks requiring precise, stable, and extended interaction through on-palm DH input, while also enabling rapid mode switching through NDH input to support efficient transitions across interaction needs. 
\section{RELATED WORK}

\subsection{Window Interactions in VR}
The rapid development and growing affordability of VR technology have opened new opportunities for productivity work~\cite{fujita2023human, simeone2023everyday}. Unlike traditional desktop setups, VR headsets are portable and body-centric, enabling users to organize and interact with multiple applications within immersive, private, and extensible virtual environments anytime and anywhere~\cite{ofek2020practical}. Commercial headsets from Meta and Apple, such as Meta Quest 3 and Apple Vision Pro, already support flexible and portable workspaces that extend beyond the constraints of physical displays, offering users large and unlimited virtual screens. Despite VR’s spatial affordances, 2D interfaces remain popular because they provide familiar layouts, efficient information organization, and cognitively lightweight interaction for detail-oriented work across diverse domains, including information tasks and creative design~\cite{8617763, 10.1145/3313831.3376652}. 

While controllers and in-air gestures are prevalent modalities to interact with virtual windows, they often lack the stability and precision required for productivity work~\cite{10.1145/3025453.3025474, simeone2023everyday}. Similarly, gaze-based interaction enables rapid navigation across windows but primarily supports discrete input, making it less suitable for continuous manipulation of virtual content~\cite{hirzle2020survey}. Therefore, researchers have integrated conventional input tools such as mice and keyboards~\cite{8943608, 10.1145/3173574.3173919}, as well as stylus and tablets~\cite{9212653, 9284682} into VR to leverage users' familiarity and capabilities. However, such input devices designed for static physical displays may have problems adapting to varying virtual displays~\cite{simeone2023everyday} and are not always portable for users~\cite{ofek2020practical}. Overall, enabling efficient interaction with large, multi-window environments without dedicated hardware remains a critical challenge for the on-the-go use of mobile VR.

\subsection{Bare-Hand Interaction}
Bare-hand interaction offers a compelling alternative for mobile VR by eliminating the need for external devices while providing a more holistic experience that combines proprioception with natural manual interactions~\cite{10.1145/3332165.3347942, 10.1145/3491102.3501898}.

\subsubsection{Asymmetric Bimanual Interaction}
When interacting with an object, users typically coordinate both hands in distinct roles, where the non-dominant hand holds the object for the dominant hand to interact with it. Guiard’s Kinematic Chain model~\cite{guiard1987asymmetric} explains that the non-dominant hand establishes the spatial frame and interaction context, while the dominant hand executes fine-grained actions relative to that established reference. This coordination leverages the benefits from human proprioception, as users can easily perceive the relative position of their hands~\cite{10.1145/258549.258571, hinckley1998two}. Such asymmetric bimanual interactions are essential for enhancing task performance of productivity tasks, particularly in terms of precision and speed~\cite{kabbash1994two, hinckley1997attention, role1999balakrishnan}.

Standard touch interfaces often struggle to support asymmetric bimanual interaction because they lack interaction context, such as the ability to distinguish between the dominant and non-dominant hands, leading to fundamental role ambiguities~\cite{webb2016wearables}. While VR headsets, such as the Meta Quest 3 and Apple Vision Pro, can track both hands simultaneously and eliminate these ambiguities, these systems typically assign both hands identical functions, overlooking the dexterity and specialized manual roles of the two hands. 

\subsubsection{Haptic Support for High Performance}
Another factor that limits the usability of bare-hand interaction for productivity work in VR is insufficient stability and precision, as well as the difficulty of controlling movements without physical support. These limitations can result in uncertain touch registration and reduced efficiency~\cite{lindeman1999towards, 10.1145/3025453.3025474, 9995404}. Prior work has addressed these challenges by introducing haptic feedback through physical surfaces~\cite{10.1145/3290605.3300243, 9995404} or customized wearable haptic devices~\cite{feuchtner2018ownershift}. A lighter-weight alternative is self-haptics, which leverages the user’s own body as a tactile medium~\cite{10.1145/3472749.3474810, 10.1145/3491102.3501898}. Hands are inherently available, low-cost, and seamlessly integrated with the human sensorimotor system through proprioception. Prior research has also demonstrated the versatility of the hand as an input surface in VR. For example, the palm has been explored as a touchscreen-like surface~\cite{10.1145/2047196.2047255, Zhang2019ActiTouch} and as a touchpad-like interface~\cite{10.1145/2325616.2325623}, enabling users to focus on virtual content while benefiting from self-haptic feedback on the skin. In addition, touch interaction on the hand surface can leverage its physical constraints, mitigating accuracy issues caused by depth perception challenges in the 3D space.

Despite these advances, reliable touch detection on the hand without external sensing hardware remains challenging. Moreover, leveraging hand-based interfaces for productivity tasks that involve large displays, multiple windows, and prolonged engagement introduces additional constraints. The small on-hand interaction space, potential instability of the hand, and ergonomic concerns during extended use, continue to hinder efficient and sustainable interaction.  

\subsection{Input Remapping for Ergonomic Interaction}
The ``gorilla-arm effect'' is a well-known limitation of bare-hand interaction in VR~\cite{hansberger2017dispelling}. Although direct hand manipulation is intuitive and widely adopted, it often requires users to extend their arms to reach virtual windows positioned high or at a distance, which is common in productivity-oriented VR environments. Input remapping offers a solution by decoupling the physical motor of the user’s hands from the virtual interaction space~\cite{wentzel2020improving, iqbal2021reducing}. This approach enables users to perform gestures at a lower, more relaxed position while their inputs are spatially mapped to eye-level or distant virtual windows. Such techniques leverage proprioception to preserve a sense of direct manipulation, even when physical and virtual hand positions are not spatially aligned. For example, ARPads~\cite{9284722}, Virtual Pads~\cite{4142852} and Redirected Pinch~\cite{ying2026redirected} introduce virtual interaction planes near the user’s body, allowing hand movements on the plane to indirectly control a cursor on a 2D window. Ownershift similarly allows users to gradually lower their hands from elevated window positions to achieve more ergonomic interaction~\cite{feuchtner2018ownershift}. 
\section{HANDPAD DESIGN}
HandPad is a bare-hand interface that leverages the user’s NDH to establish a mobile spatial frame and interaction context. Defined relative to that established reference, the DH performs touch interactions on the NDH with self-haptic feedback or mid-air gestures to modulate the frame. Through asymmetric bimanual coordination, the NDH further employs distinct gestures to enable seamless mode switching. Overall, HandPad supports a suite of interaction techniques that operate across three levels of components: the interaction window, which defines the spatial frame; the virtual content within the window; and the application window.

\subsection{Design Goal}
While bare-hand mid-air interaction is widely adopted in commercial VR headsets due to its intuitiveness and availability, interacting with hand interfaces can introduce several challenges. Here, we present our design goals to ensure HandPad interactions can effectively be used for productivity work in VR. 

\paragraph{\textit{Efficient}}
Previous work has shown the benefit of bimanual asymmetry~\cite{guiard1987asymmetric, kabbash1994two, hinckley1997attention, role1999balakrishnan} and haptic feedback~\cite{10.1145/3472749.3474810, 10.1145/3025453.3025474, 10.1145/3290605.3300243} for enhancing the interaction efficiency in VR. We assigned different roles to the two hands. The NDH serves as (1) a spatial frame to establish the interaction window and provide a reference to anchor the DH's motions; (2) a tangible surface stabilizing the frame to support precise DH touch interactions on the NDH with self-haptics; (3) a kinesthetic mode-switching modality using gesture to phrase interactions and create context-specific meanings for different inputs such as manipulating contents or windows. 

\paragraph{\textit{Comfortable}}
Directly interacting with virtual content using hands in VR may cause fatigue and discomfort for users due to the gorilla-arm effect~\cite{hansberger2017dispelling}. Especially for prolonged productivity tasks, users need to perform interactions with a comfortable posture. Inspired by previous work that created a virtual touchpad and allowed users to interact with it at a lower and more comfortable location~\cite{9284722, 4142852}, we employ the hand as a touchpad. Users can perform gestures at a lower and more comfortable posture. Meanwhile, both hands and their interactions will be remapped to the corresponding virtual window, creating the sensation that the user is directly touching the content.

\paragraph{\textit{Stable}}
Another limitation of hand-based interfaces in VR is users’ limited ability to maintain stable hand positions in mid-air, particularly during 2D window interaction. Even minor tremors can degrade performance in simple tasks such as selection. To address this issue, we introduce a snapping mechanism that anchors the interaction window to the application window, thereby stabilizing the spatial reference frame. This allows the dominant hand (DH) to perform touch interactions or mid-air gestures relative to a fixed frame, reducing the impact of unintended hand jitter.

\paragraph{\textit{Complementary}}
Designed gestures should not conflict with users’ natural resting postures or commonly adopted mid-air gestures, such as pinching and grabbing in current commercial VR headsets. We therefore assign uncommon yet easy-to-perform hand postures to the NDH for specific functions, such as mode switching, while preserving familiar DH gestures, including pointing, pinching, and multi-touch input.

\paragraph{\textit{Versatile}}
Beyond basic interactions such as clicking, dragging, and scrolling, hand-based interfaces must support the complex and compound manipulations inherent to productivity work. Productivity tasks frequently require coordinated multi-touch operations, including zooming and rotating content, transferring information across multiple windows, and managing application windows through moving, resizing, and rotating.  

\paragraph{\textit{Generalizable}}
Although the physical hand provides only a limited interaction surface, it must flexibly accommodate application windows of varying sizes. As proposed, the hand serves as a spatial reference frame that instantiates an interaction window on top of application windows. This interaction window should be resizable and seamlessly movable across multiple applications through coordinated hand gestures, enabling fluid transitions between contexts. Furthermore, when users need to interact with content beyond the bounds of the interaction window, the hand can function as a stable anchor that supports an effectively unbounded virtual interaction space around it. This allows users to extend their input beyond the physical hand surface and perform mid-air interactions, such as clicking, while preserving spatial continuity and ergonomic comfort.

\subsection{Design Space}
We present a design space for high-performance bimanual interaction (Table~\ref{tab:design-space}). The rows represent distinct functional roles of NDH input and the columns differentiate the input space.

\definecolor{headergray}{RGB}{242,242,242}
\definecolor{handpadgray}{RGB}{230,230,230}

\begin{table}[t]
\centering
\caption{Design Space of Bimanual Interaction Techniques.}
\label{tab:design-space}
\footnotesize
\setlength{\tabcolsep}{4pt}
\renewcommand{\arraystretch}{1.18}

\begin{tabularx}{\columnwidth}{
>{\raggedright\arraybackslash}m{1.9cm}
>{\centering\arraybackslash}m{1cm}
>{\centering\arraybackslash}m{2.5cm}
>{\centering\arraybackslash}m{2cm}}
\toprule
\rowcolor{headergray}
\textbf{Dimension} & \textbf{HandPad} & \textbf{Hand/Arm} & \textbf{Device} \\
\midrule

D1 Frame of Reference
& \ding{51}
& Imaginary interfaces \cite{10.1145/1866029.1866033}, PalmRC \cite{10.1145/2325616.2325623}
& VXSlate \cite{10.1145/3461778.3462076}, BiShare \cite{10.1145/3313831.3376233} \\ \midrule

D2 Changing Modes
& \ding{51}
& Hand interfaces \cite{10.1145/3491102.3501898}, BlyncSync \cite{10.1145/3313831.3376132}
& Body-Centric Interaction \cite{10.1145/2371574.2371599} \\ \midrule

D3 Creating Shortcuts
& \ding{51}
& Hand interfaces \cite{10.1145/3491102.3501898}, Hand Proximate \cite{10.1145/3563657.3596117}
& Body-Centric Interaction \cite{10.1145/2371574.2371599} \\ \midrule

D4 Constraining Movement
& \ding{51}
& Self-haptics \cite{10.1145/3472749.3474810}, Plane, Ray, Point \cite{10.1145/3332165.3347916}
& BiShare \cite{10.1145/3313831.3376233}, VXSlate \cite{10.1145/3461778.3462076} \\ \midrule

D5 Controlling Cursor Behavior
& \ding{51}
& Gaze-Shifting \cite{pfeuffer2015gaze}, Pen + Touch \cite{913781}
& VXSlate \cite{10.1145/3461778.3462076}, Arc-pad \cite{10.1145/1622176.1622205} \\

\bottomrule
\end{tabularx} 
\vspace{-4mm}
\end{table}

\paragraph{\textit{D1 Establishing Spatial Frame of Reference}} 
The hand itself can serve as a spatial frame of reference for interaction. For example, Imaginary Interfaces~\cite{10.1145/1866029.1866033} allow users to define an interaction space by forming an L-shape with the NDH and performing drawing gestures with the DH within the defined frame. PalmRC~\cite{10.1145/2325616.2325623} similarly leverages the user’s palm as an interactive surface, enabling menu navigation through bare-hand input while maintaining visual focus on the display. Other systems establish interaction frames using external devices, such as Bishare~\cite{10.1145/3313831.3376233}, VXSlate~\cite{10.1145/3461778.3462076} and Breaking the Screens~\cite{9212653}, which rely on smartphones or tablets to support large-window or multi-window manipulation. While effective, such approaches depend on dedicated hardware. In contrast, our work focuses on bare-hand interaction, using the non-dominant hand to establish and manipulate the interaction frame directly within VR.

\begin{figure*}[!htbp]
 \centering
 \includegraphics[width=\linewidth]{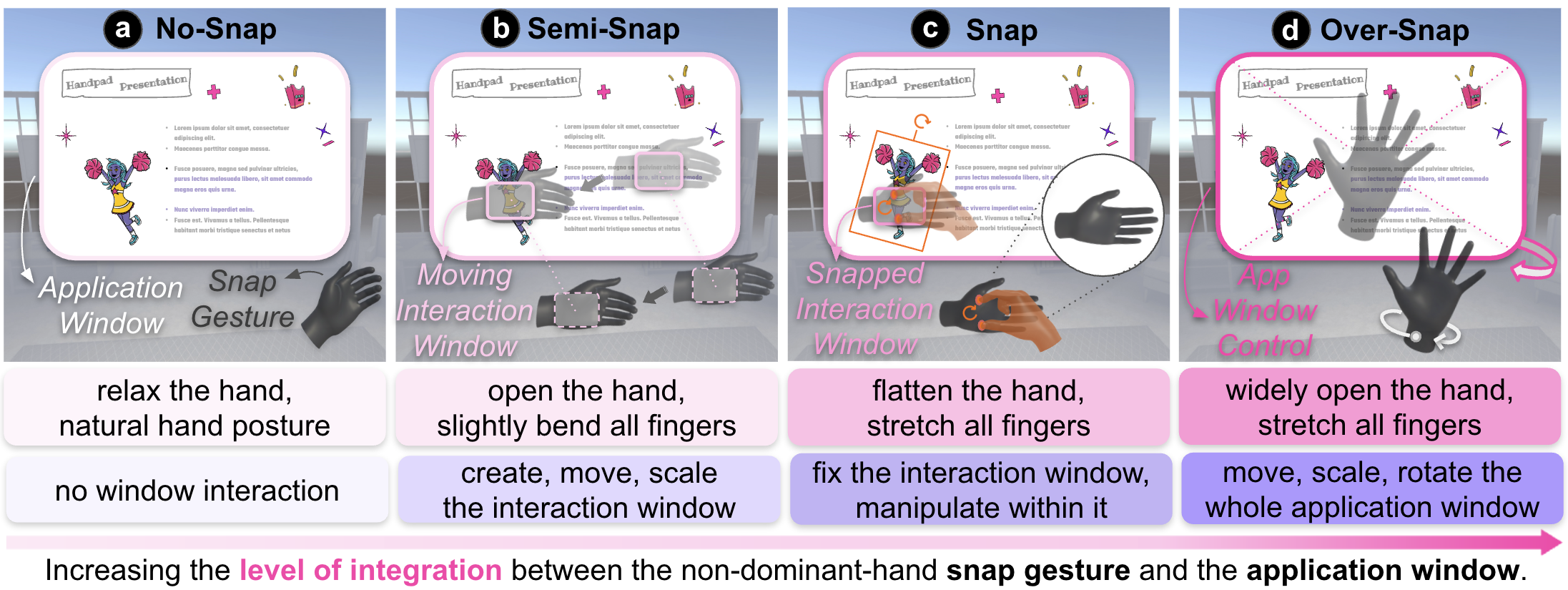}
 \vspace{-5mm}
 \caption{Non-dominant hand gestures set different interaction modes. (a) No-Snap: Natural hands, no window interaction. (b) Semi-Snap: Bend the NDH to create, move, and scale the interaction window. (c) Snap: Fix the interaction window and manipulate 2D content via DH gestures on the flat NDH. (d) Over-Snap: Use a widely open NDH to move, scale, and rotate the entire application window. Dark gray hands indicate the physical NDH; orange hands indicate the physical DH; translucent hands indicate the remapped hands on the window.  \vspace{-2mm}}
 \label{fig:snap_gesture}
\end{figure*}

\paragraph{\textit{D2-3 Mode Switching and Gesture-Based Shortcuts}}

Gestures provide an efficient mechanism for rapid mode switching and the creation of interaction shortcuts. Prior work has demonstrated that bimanual gestures can support seamless transitions between interaction states, such as simulating virtual objects to encode different modes~\cite{10.1145/3491102.3501898}. Touch-based gestures on wearable devices have similarly enabled multi-modal interaction and shortcut activation~\cite{10.1145/3313831.3376132}. Related research in body-centric interaction further shows that anchoring input to specific body parts, such as the wrist, can trigger specialized functions and shortcuts~\cite{10.1145/2371574.2371599}. Importantly, the hand is particularly well-suited for mode switching because it provides kinesthetic feedback and proprioception cues, allowing users to maintain embodied awareness of interaction states and reducing mode ambiguity~\cite{surale2019experimental, 10.1145/3491102.3501898}. 

\paragraph{\textit{D4 Constraining Movements}} 
Constraining physical movement is an effective strategy for improving interaction accuracy in VR~\cite{10.1145/3267782.3267788}. Prior work has shown that interacting on physical surfaces can restrict unintended motion and enhance control compared to purely in-air gestures~\cite{10.1145/3025453.3025474}. For example, holding a device while performing gestures introduces passive mechanical constraints that stabilize input~\cite{10.1145/3313831.3376233, 10.1145/3461778.3462076}. Additionally, one hand can provide tactile guidance to the other, enabling self-haptic feedback that constrains and refines movement~\cite{10.1145/3472749.3474810}. Symbolic gesture constraints, such as those introduced in Plane, Ray, and Point~\cite{10.1145/3332165.3347916}, further demonstrate how constrained input can improve precision.

\paragraph{\textit{D5 Controlling Cursor Behavior}} 
Precise cursor control and efficient target switching are central challenges in large or multi-window environments. Prior work has explored various control-to-display (C-D) gain strategies, including stitching and pointer warping, to bridge spatial gaps between displays and facilitate efficient selection and window switching~\cite{10.1145/1868914.1869036,10.5555/1992917.1992939}. Similarly, techniques have been proposed to map a small input space, such as a touchscreen, to a large output space, such as a wall-sized display, by dynamically adjusting cursor movement and gain~\cite{10.1145/2470654.2470773,10.1145/1622176.1622205}. Bimanual input has also been leveraged to modulate cursor behavior. For example, Brandl et al.~\cite{10.1145/1385569.1385595} combine pen input in the DH with touch input in the NDH to support coordinated interaction. Gaze-shifting~\cite{pfeuffer2015gaze} further integrates gaze as contextual input alongside pen and touch, demonstrating how multiple modalities can dynamically influence cursor positioning and selection.

\subsection{HandPad Interaction}
HandPad is designed to address the interaction dimensions outlined above. It employs the user’s NDH to establish a mobile spatial frame and interaction context, enabling fluid mode switching while allowing the DH to perform fine-grained manipulations relative to this reference frame. 

\subsubsection{Non-Dominant Hand Mode Gestures}
HandPad enables the NDH to interact with the application window through four progressively window-integrated modes: (1) \textbf{No-Snap}, (2) \textbf{Semi-Snap}, (3) \textbf{Snap}, and (4) \textbf{Over-Snap} as shown in Figure~\ref{fig:snap_gesture}. Users transition between modes by performing distinct NDH gestures. Specifically, changes in finger flexion angles and inter-finger spreading angles are used to differentiate modes, and increasing finger stretchiness corresponds to a higher level of integration with the application window. 

In No-Snap (Figure~\ref{fig:snap_gesture}a), the NDH remains relaxed with minimal finger extension, resulting in no window interaction. In Semi-Snap (Figure~\ref{fig:snap_gesture}b), users slightly bend the NDH to create, move, and scale the interaction window, thereby establishing the interaction frame. When users stretch their fingers to form a flat surface, the system enters Snap mode (Figure~\ref{fig:snap_gesture}c), which fixes the interaction window and enables manipulation of 2D content via DH gestures on the flat NDH. Finally, in Over-Snap (Figure~\ref{fig:snap_gesture}d), users fully spread their fingers, triggering the highest level of window-integration and allowing manipulation of the entire application window, including moving, scaling, and rotating.

\begin{table}[t]
\centering
\small
\setlength{\tabcolsep}{4pt}
\caption{Joint configurations across different NDH snap modes.}
\label{tab:joint-angles}
\begin{tabular}{p{3.4cm}cccc}
\toprule
 & No-Snap & Semi-Snap & Snap & Over-Snap \\
\midrule
\raisebox{-0.4\height}{\includegraphics[height=0.7cm]{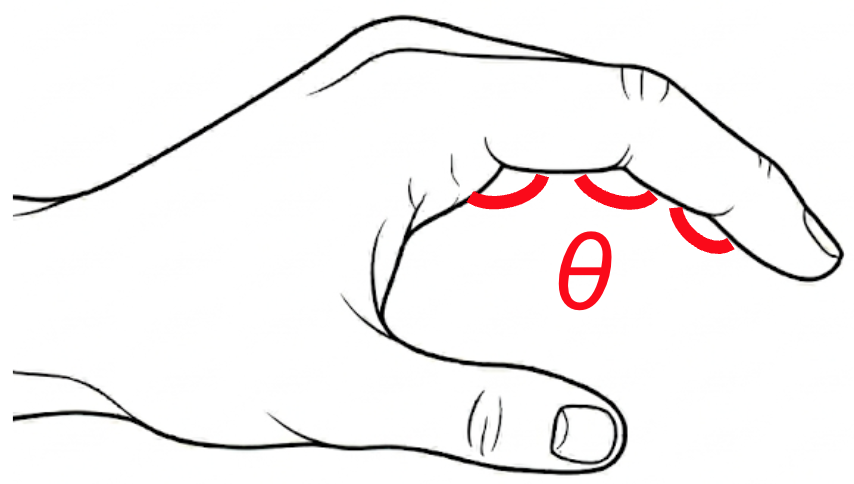}}
\hspace{4pt} Joint flexion angle 
& $200$--$250^\circ$ & $80$--$150^\circ$ & $40$--$60^\circ$ & $<40^\circ$ \\

\raisebox{-0.4\height}{\includegraphics[height=0.9cm]{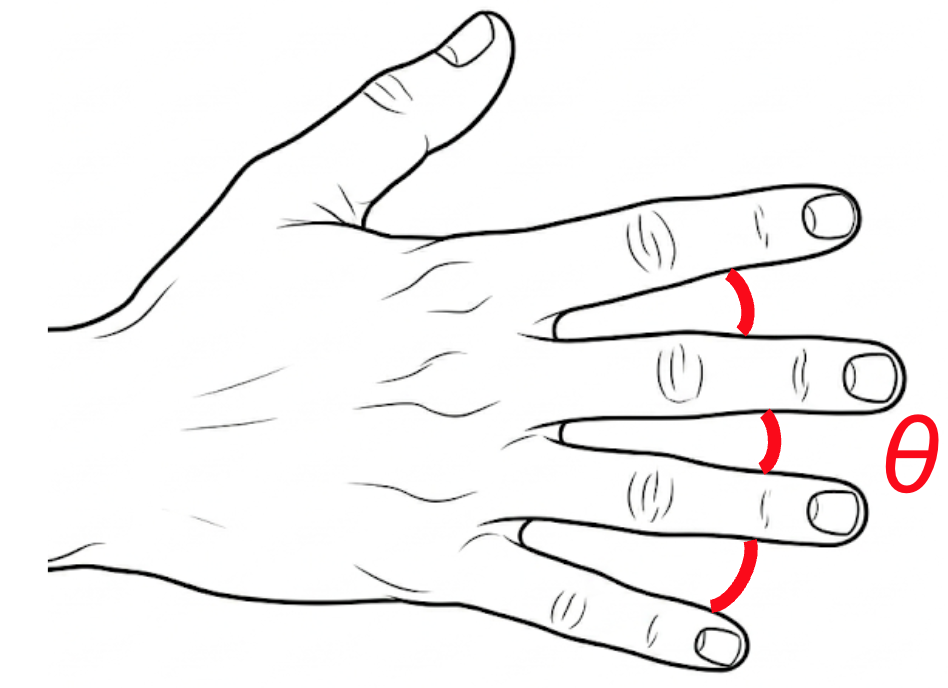}}
\hspace{4pt} Inter-finger angle
& $30$--$40^\circ$ & $30$--$40^\circ$ & $30$--$40^\circ$ & $>60^\circ$ \\
\bottomrule
\end{tabular}
\vspace{-5mm}
\end{table}

To ensure that NDH gestures do not conflict with users’ natural resting postures or commonly adopted mid-air gestures, such as pinching and grabbing in commercial VR headsets, we conducted an informal study with four participants. Participants were asked to rest their hands naturally and perform common mid-air gestures while we recorded joint flexion and inter-finger angles. We observed that, during these activities, the total joint flexion angle exceeded $250^\circ$, while the total inter-finger angle remained below $30^\circ$.

To minimize interference with everyday gestures, we designed HandPad NDH gestures to occupy a distinct range of hand configurations, with total joint flexion angles (JFA) between $40^\circ$ and $250^\circ$ and total inter-finger angles (IFA) between $30^\circ$ and $60^\circ$. As summarized in Table~\ref{tab:joint-angles}, No-Snap occurs when JFA is within $150^\circ$–$250^\circ$ and IFA is within $30^\circ$–$40^\circ$. Semi-Snap is triggered when JFA decreases to $80^\circ$–$150^\circ$. Snap occurs when JFA further decreases to $40^\circ$–$60^\circ$. Finally, Over-Snap is activated when the hand becomes widely opened, characterized by JFA $<40^\circ$ and IFA $>60^\circ$.

\begin{figure}[!t]
 \centering
 \includegraphics[width=\columnwidth]{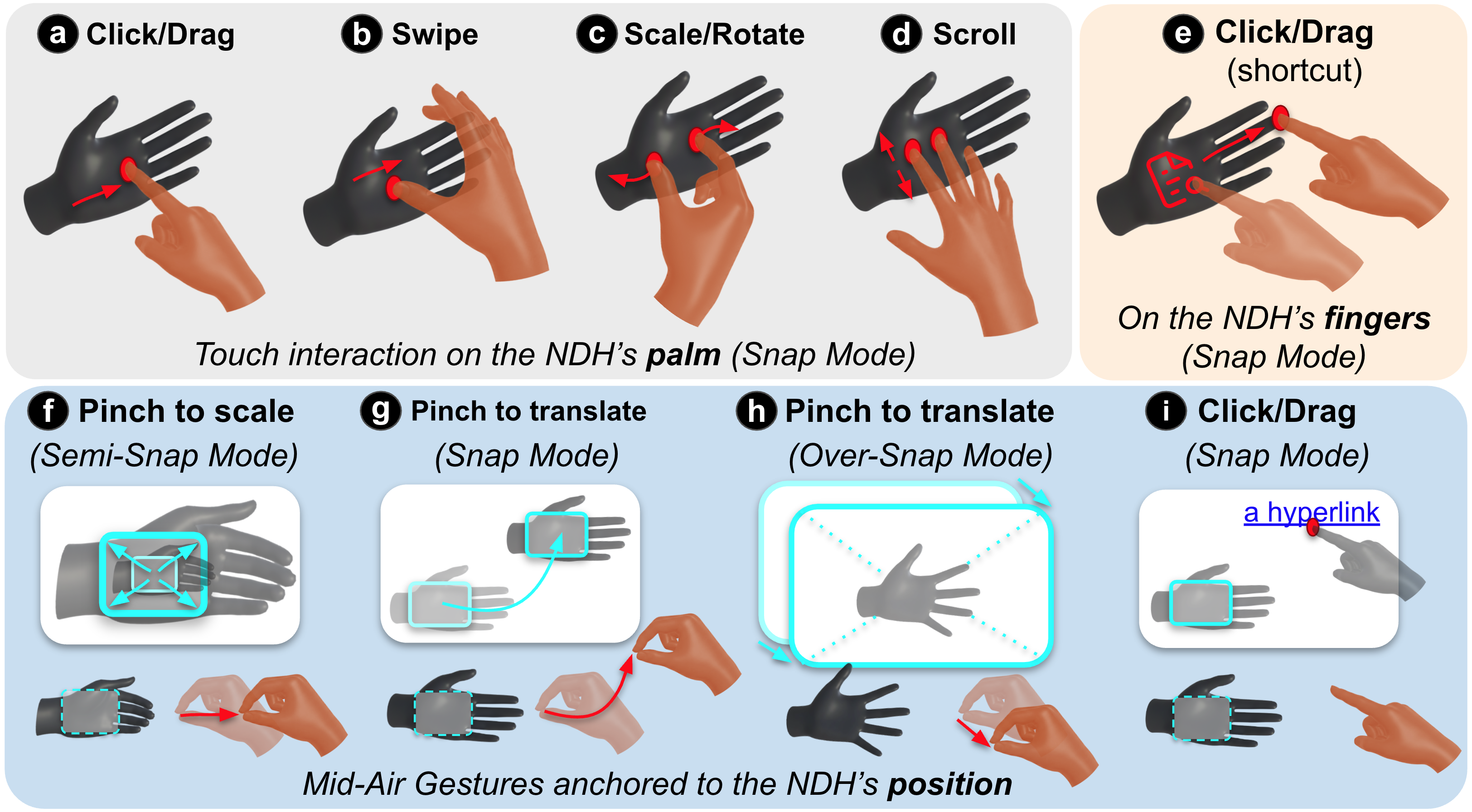}
 \vspace{-4mm}
 \caption{Dominant-hand manipulation gestures. Snap mode enables palm-based touch interactions for (a) click/drag, (b) swipe navigation, (c) scale/rotate, and (d) scroll; finger touch supports (e) shortcut operations. NDH-anchored mid-air gestures include (f) pinch to scale (Semi-Snap), (g) pinch to translate the interaction window (Snap), (h) pinch to translate the application window (Over-Snap), and (i) click/drag when content lies beyond the hand surface. \vspace{-3mm}}
 \label{fig:manipulation_gesture}
\end{figure}

\subsubsection{Dominant Hand Manipulation Gestures}

HandPad leverages the DH to perform touch gestures on the NDH palm, supporting both single-touch and multi-touch interactions, as well as mid-air gestures anchored to the NDH’s spatial position for diverse window manipulations (Figure~\ref{fig:manipulation_gesture}). In the following, we describe the touch detection method used to identify DH–NDH contact and the gesture recognition mechanism implemented through a state-machine-based approach.

\paragraph{\textit{Touch Detection}}
We developed a vision-based touch detection model to identify both single-point and multi-point contact on the NDH surface using the built-in hand tracking of the Quest headset. The primary feature for touch detection is the distance between the DH fingertip and the NDH palm surface. However, a fixed distance threshold proved unreliable due to the uneven geometry of the hand, which caused missed touches in some regions and premature detections in others. To address this, we trained a random forest classifier to adaptively determine touch based on fingertip distance and spatial location on the hand surface. To reduce tracking jitter, we applied a 1€ filter~\cite{casiez20121} to hand and finger motion, using parameters $f_{c\ min}= 0.9$ and $\beta = 90$.

We collected 5,000 touch samples by tapping the NDH palm with the DH index fingertip across different regions, and 5,000 no-touch samples by hovering above the same areas. Each sample included the fingertip-to-palm distance and the normalized $(x, y)$ coordinates of the ray-cast intersection on the palm surface. The dataset was split evenly for training and evaluation. We trained a random forest classifier with $N = 100$ trees and a maximum depth of 2. The initial model achieved an accuracy of 89.3\%, with false negatives of 3.2\% and false positives of 7.5\%. Given the limitations of vision-based tracking and potential occlusion, a small number of false positives are tolerable, as near-surface detections typically precede actual contact \cite{10.1145/2047196.2047255}. However, false negatives result in missed touch events and are more detrimental to interaction continuity. We therefore tuned the confidence threshold to reduce false negatives to 1.3\%, accepting an increase in false positives to 12.2\%.
During HandPad interaction, the trained model runs in real time to predict the touch state of each DH fingertip. 

\paragraph{\textit{Manipulation Gesture Recognition}}

\begin{figure}[!t]
 \centering
 \includegraphics[width=\columnwidth]{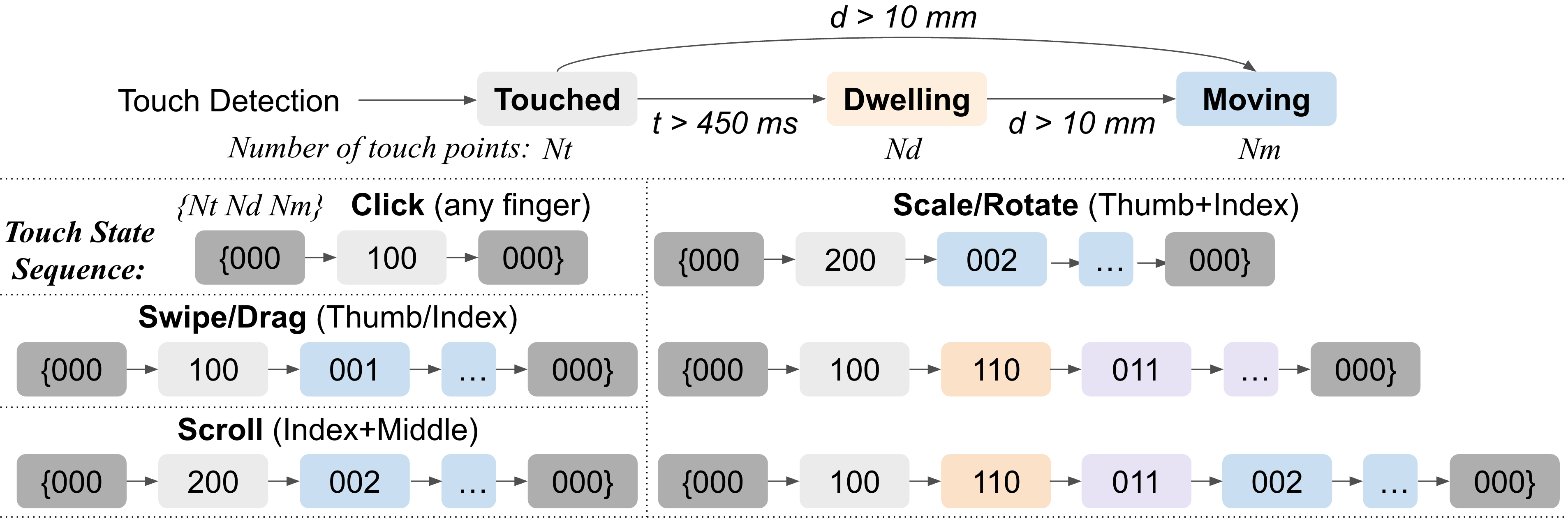}
 \vspace{-5mm}
 \caption{Touch state representation and gesture sequences. Each touch point can be in one of three states: touched, dwelling, or moving. Each state is encoded by the number of points in these three phases. Example state sequences are shown for four gestures: click, scroll, swipe/drag, and scale/rotate. \vspace{-4mm}}
 \label{fig:touchstate}
\end{figure}

Once contact between a DH manipulation fingertip and the NDH palm surface is detected, the touch is classified into three states: touched, dwelling, and moving, as shown in Figure~\ref{fig:touchstate}, following prior touchpad-based approaches~\cite{10.1145/2858036.2858452}. 
Specifically, a touch enters the dwelling state if the contact point remains within the same location for longer than 450 ms. A touch or dwelling enters the moving state if the displacement of the contact point exceeds 10 mm. The touch state of the manipulation fingertip at a given moment is represented as three consecutive numbers: the number of touch points ($N_t$), the number of dwelling points ($N_d$), and the number of moving points ($N_m$). A gesture is then recognized as a sequence of touch states $\{N_t N_d N_m\}$. Unlike traditional touchpads, which cannot easily distinguish individual fingers and may therefore produce gesture ambiguities~\cite{10.1145/2858036.2858452}, our system integrates touch-state sequences with the identities of specific DH fingers. This allows us to differentiate gestures performed on the NDH palm surface. As illustrated in Figure~\ref{fig:touchstate}, the sequence $\{000,100,000\}$ represents a click gesture. The sequence $\{000,100,001\}$ represents a swipe gesture when performed by the thumb, or a drag gesture when performed by the index finger. The sequence $\{000,200,002\}$ represents a scale or rotate gesture when performed by the thumb and index finger, and a scroll gesture when performed by the index and middle fingers. In practice, scale and rotate interactions may produce additional state sequences if one finger remains in the dwelling state while the other moves. 

Mid-air gesture detection follows the interaction techniques used in the Oculus Quest Pro system. Direct click and drag gestures are detected when the manipulation fingertip collides with the virtual window. Pinch gestures are recognized based on the distance between the thumb and index fingertips.

\subsection{Interaction Techniques}
By combining NDH-enabled interaction modes (Figure~\ref{fig:snap_gesture}) with DH manipulation gestures (Figure~\ref{fig:manipulation_gesture}), we establish a spatial reference frame using an interaction window created by the NDH, allowing both hands to perform manipulations relative to this frame. Based on this design, we develop interaction techniques that operate at four levels: (1) across-window interaction (in Semi-Snap), (2) interaction on a window (in Snap), and (3) window management (in Over-Snap), reflecting an increasing level of integration between the NDH and the application window. We also discuss mid-air gestures that complement these interactions. 

\subsubsection{Interaction Window}
The interaction window defines the active interactive area on the application window. It is a floating window created and manipulated under the \textbf{Semi-Snap} mode. Users create the interaction window by gazing at or pointing the NDH toward the desired location while performing the Semi-Snap gesture (Figure~\ref{fig:snap_gesture}b). To accommodate large application windows and support seamless transitions, the interaction window can be relocated and resized dynamically. 

Users can relocate the interaction window by gazing at or pointing to another location while remaining in Semi-Snap mode. If \textbf{Snap} mode has already been triggered, users can also reposition the interaction window by performing a pinch gesture with the DH and dragging it around the NDH, as shown in Figure~\ref{fig:snap_gesture}g.

To resize the interaction window, users can perform a pinch gesture with the DH (Figure~\ref{fig:manipulation_gesture}f). Moving the pinched fingers toward the NDH reduces the size of the interaction window, while moving them away enlarges it. The size of the interaction window influences cursor behavior on the application window because interactions on the NDH are mapped one-to-one to the interaction window. A larger interaction window produces a higher control-to-display (C/D) gain, enabling faster navigation and large-scale manipulations such as scrolling quickly or interacting with large images. Conversely, a smaller interaction window supports finer control for tasks such as signing or annotation.

\subsubsection{Across-window Interaction}

\begin{figure}[tb]
 \centering
 \includegraphics[width=\columnwidth]{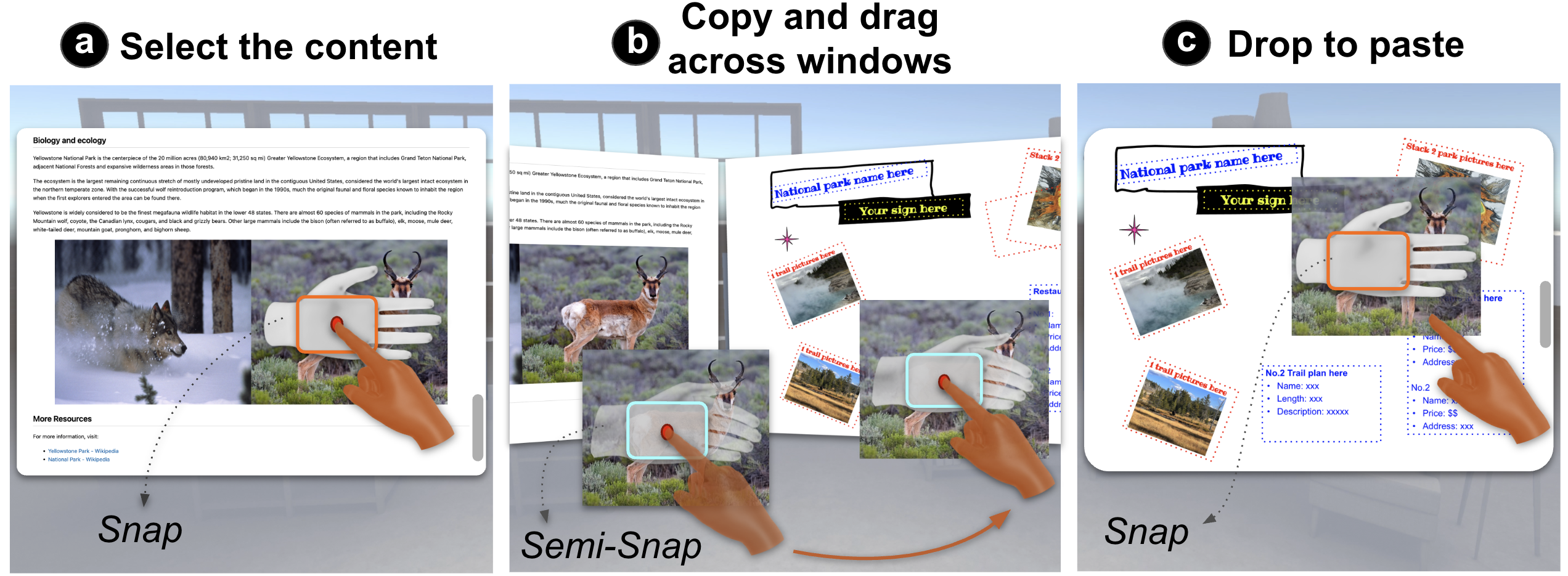}
 \vspace{-4mm}
 \caption{Across-window interaction example: drag-and-drop to copy-paste. White hands indicate the virtual NDH and orange hands indicate the virtual DH.\vspace{-3mm}}
 \label{fig:draganddrop}
\end{figure}

During \textbf{Semi-Snap} (Figure~\ref{fig:snap_gesture}b), the interaction window can be translated, which is particularly useful for large-window and across-window interactions, where long-distance manipulations on the NDH palm would otherwise require repeated small drags. For example, to move an image from the bottom of an application window to the top, or even across different application windows (Figure~\ref{fig:draganddrop}), the user can long-press on the NDH with the DH to select the image. The selected image is then attached to the interaction window and moves with it as the NDH relocates the window to the desired location. The user completes the operation by lifting the DH fingertip from the NDH and dropping the image at the target location.

\subsubsection{Interaction on a Window}
When users interact with 2D content on an application window (e.g., text, images, or buttons), the interaction window can be fixed over the content using the \textbf{Snap} gesture, as shown in Figure~\ref{fig:snap_gesture}c. Once the interaction window is stabilized, users can relax the NDH into a more comfortable posture while performing fine-grained interactions on it with the DH. Based on the interaction contexts established by the NDH during \textbf{Snap} mode, three types of interactions are enabled: (1) touch interactions on the NDH palm, (2) touch interactions on the NDH fingers, and (3) mid-air gestures anchored to the NDH position.

\paragraph{\textit{Touch interactions on the NDH palm}}
While controllers and mid-air gestures support only a limited set of interactions due to the lack of physical surface support and the under-utilization of finger dexterity, HandPad leverages the tangible flat surface of the palm while allowing both hands to move freely. Because the palm area is comparable in size to a touchpad, HandPad supports classic single-touch and multi-touch interactions similar to those on a touchpad. 

For example, users can repeatedly click and drag on the NDH palm with any DH finger to control a cursor or perform stroke-based input (Figure~\ref{fig:manipulation_gesture}a). To navigate across a webpage, users can swipe the DH thumb on the NDH palm to move up, down, left, or right (Figure~\ref{fig:manipulation_gesture}b). For precise manipulation of 2D content, such as scaling or rotating text or images, users can use the DH thumb and index finger to perform fine multi-touch adjustments on the NDH palm (Figure~\ref{fig:manipulation_gesture}c). To scroll through a webpage, users can perform brief two-finger swipes with the DH index and middle fingers on the NDH palm (Figure~\ref{fig:manipulation_gesture}d). These touch interactions operate under relative control and incorporate inertial effects based on the speed of finger movement. For instance, faster swipes on the NDH palm produce longer scrolling distances, which is particularly beneficial when navigating large windows.

\paragraph{\textit{Touch interactions on the NDH fingers}}

\begin{figure}[tb]
 \centering
 \includegraphics[width=\columnwidth]{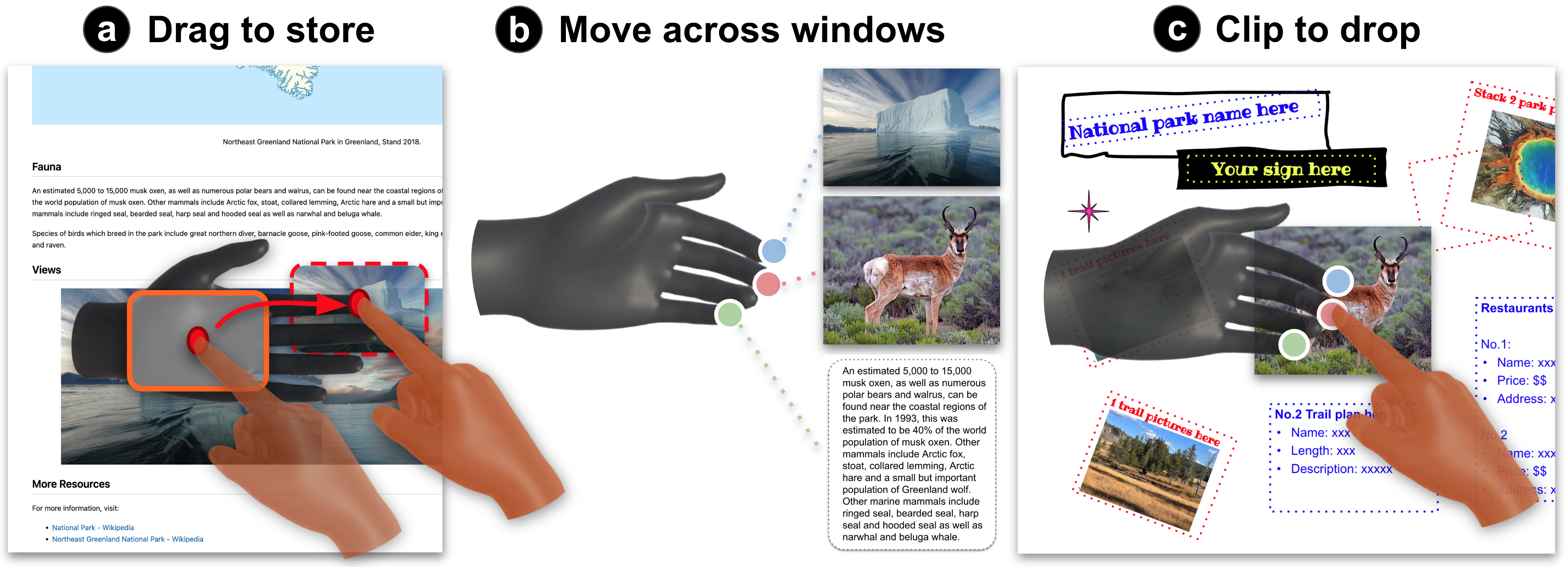}
 \vspace{-4mm}
 \caption{Example of touch interactions on the NDH fingers: shortcuts for operating a temporary clipboard. \vspace{-3mm}}
 \label{fig:shortcut}
\end{figure}

Beyond the NDH palm surface, fingers can also be leveraged for efficient touch interactions, such as shortcuts~\cite{913781}, since users can easily locate each finger through proprioception. As an example scenario shown in Figure~\ref{fig:manipulation_gesture}e\&Figure~\ref{fig:shortcut}, HandPad enables a temporary clipboard that allows users to quickly store multiple 2D contents (e.g., text or images) or files across different NDH fingers. Users first select the desired content on the NDH palm and then move it to a specific fingertip using a DH manipulation finger, effectively storing the item on that finger. The stored contents then move together with the NDH, allowing users to transport them across multiple application windows. To retrieve an item, users simply click on the corresponding NDH fingertip and drag it back to the NDH palm, placing the stored content at the desired target location.

\paragraph{\textit{Mid-air gestures anchored to the NDH position}}
While the NDH palm supports diverse and fine-grained touch interactions similar to a touchpad, application windows in VR can be extremely large compared to the small surface area of the hand. In scenarios where the interaction window is fixed by the NDH, users may need to perform simple, quick, and repetitive actions outside this window, such as clicking on items beyond the interaction area. Repeatedly switching NDH modes or relocating the interaction window for these minor interactions can become tedious. In contrast, mid-air gestures are well-suited for such lightweight interactions without physical surface or haptic feedback. HandPad therefore allows the NDH to serve as an anchor while enabling an unlimited virtual interaction space around the interaction window. This allows users to directly point and interact with the application window in mid-air using the DH, such as selecting hyperlinks (Figure~\ref{fig:manipulation_gesture}i).

\subsubsection{Application Window Management}
Efficient manipulation of entire application windows, such as relocating, minimizing or opening, rotating, and resizing, is essential for managing multiple windows in VR. HandPad enables these operations through the \textbf{Over-Snap} gesture performed by the NDH (Figure~\ref{fig:snap_gesture}d), which snaps the entire application window to the NDH for direct manipulation. To relocate an application window, users can point at the target application window and perform the Over-Snap gesture using NDH. The application window then attaches to the NDH, allowing users to move it to a new location before releasing the gesture to place it. However, given the large 3D virtual space, moving windows solely with the NDH may be inefficient when the target location lies beyond arm’s reach. To address this limitation, HandPad allows the DH to perform relative control using pinch gestures (Figure~\ref{fig:manipulation_gesture}h). By pinching relative to the NDH, users can translate the application window along the x, y, and z axes. Window orientation can be adjusted by rotating the NDH through wrist twisting, which allows the window to face different directions. Users can also resize the application window by performing a scale gesture with the DH thumb and index finger on the NDH during Over-Snap mode.

\subsection{Hand Remapping}
When users initiate HandPad interaction, both physical hands are translationally and rotationally remapped to the application window to enable direct-like manipulation. Since the physical NDH serves as the spatial reference frame represented by the interaction window, we retarget the physical NDH to the location of the interaction window and render a corresponding virtual NDH on the application window. The physical DH is then retargeted according to the relative position and orientation between the physical and virtual NDH (Figure~\ref{fig:snap_gesture}b–d). After hand remapping, users can relax their physical hands while the virtual hands remain aligned with the application window. 
\section{HANDPAD STUDY}
Prior research has highlighted the advantages of asymmetric bimanual interaction~\cite{10.1145/258549.258571, hinckley1998two} and hand-based interfaces that provide self-haptic feedback and tangible manipulation~\cite{10.1145/3472749.3474810, 10.1145/3491102.3501898}. Building on these insights, we conducted a study to investigate how HandPad interaction techniques support VR productivity tasks, using multi-window information and design work as representative scenarios.

\subsection{Apparatus}
Our study involved web applications displayed across multiple windows. The applications were developed and hosted on a Python Flask server. Application windows were rendered in VR using a Unity web browser plugin by Vuplex.
HandPad interactions performed on windows in VR were transmitted to the web applications through a custom event sender–listener system, enabling real-time interaction. Hand tracking, touch interaction, and mid-air gesture recognition were implemented using the Meta XR Interaction SDK in Unity. The study application ran on a laptop equipped with an Intel Core i7 processor and an NVIDIA RTX 3080 GPU.   

\subsection{Participants}
Nine participants (6 male and 3 female) were recruited from the university. Their experience with VR varied. Two participants reported frequent use (daily or weekly), while the others used VR occasionally for gaming, product exploration, or participation in research studies. Each study session lasted approximately 60 minutes, and participants received \$20 USD in compensation.

\subsection{Procedure}
Participants were seated and wore a Meta Quest Pro VR headset throughout the study. The study began with a practice session to familiarize participants with the HandPad system, including the four NDH snap modes and DH manipulation gestures for different types of window interactions: (1) interaction window manipulation, (2) interaction on a window, (3) window management, and (4) complementary mid-air gestures. Participants practiced these interactions by following the experimenter’s instructions on a tutorial window until they felt comfortable using the system. The practice session lasted approximately 5–10 minutes. Afterward, participants used HandPad in two VR scenarios: (1) an open-book quiz and (2) a poster design task, each requiring interactions across three to four windows. During the tasks, the experimenter provided verbal guidance on the next steps and suggested interactions relevant to each scenario. After completing both scenarios, participants filled out a survey evaluating their experience with each HandPad interaction technique. Finally, they completed a post-study questionnaire and participated in a five-minute semi-structured interview to provide general feedback on using HandPad for productivity work in VR.

\begin{figure}[tb]
 \centering
 \includegraphics[width=\columnwidth]{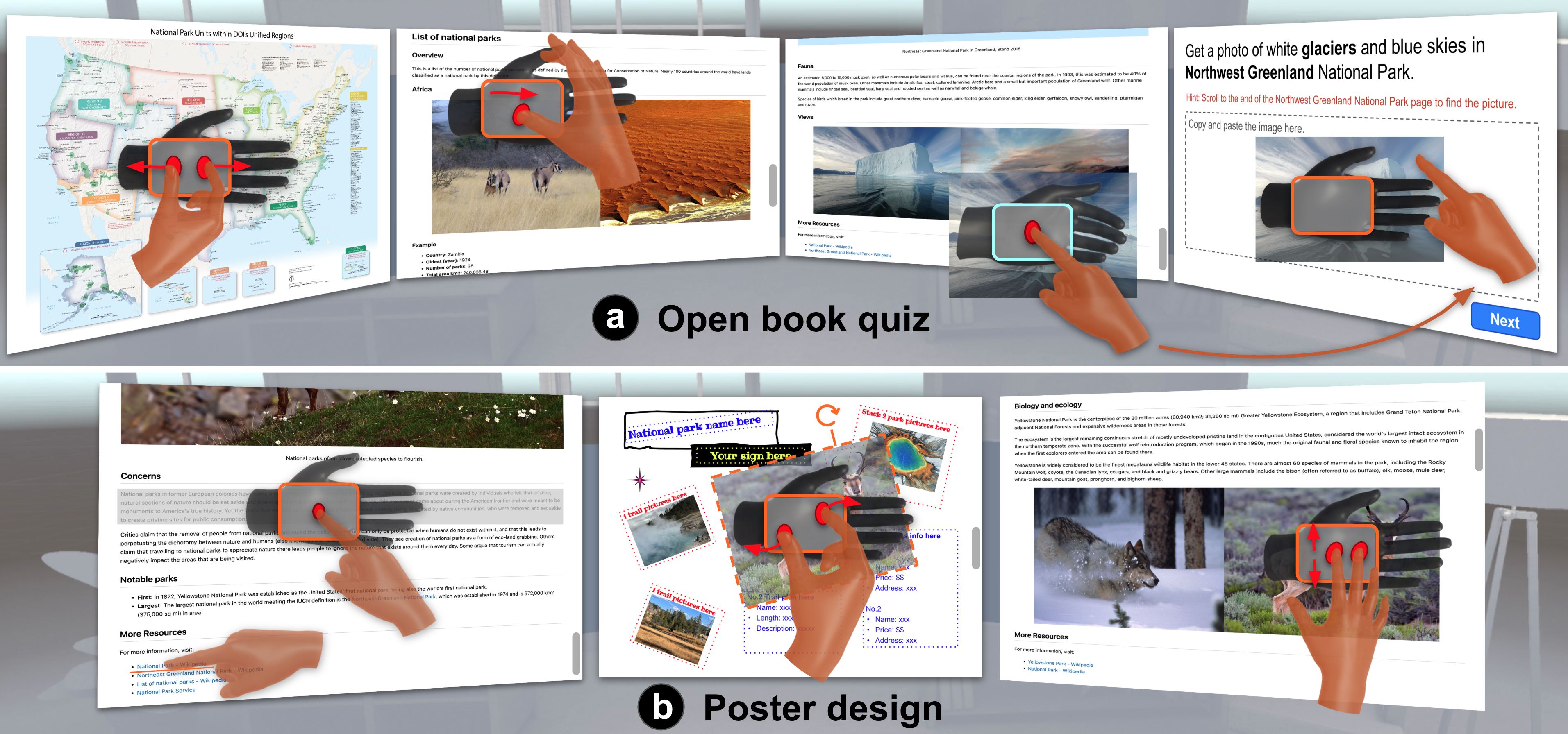}
 \caption{Screenshots of the scenarios used in the study.
 \vspace{-5mm}}
 \label{fig:study_task}
\end{figure}

\subsection{Tasks}
Participants interacted with the HandPad system in two productivity-oriented tasks: (1) completing an open-book quiz and (2) designing an illustrated poster.

\paragraph{\textit{Open Book Quiz}}
The quiz task involved four application windows, as shown in Figure~\ref{fig:study_task}a. One application window displayed the quiz questions, while the other three presented information sources from which participants could find the answers. These sources included (1) a map that participants needed to zoom and pan, and (2) two Wikipedia pages that participants navigated to locate relevant text and image information. After finding the required information, participants either clicked on the question page to enter an answer or copied content by dragging text or images from the information pages into the answer box. They then clicked a button to proceed to the next question. This quiz scenario was designed to allow participants to experience frequent information search and manipulation tasks using the HandPad interaction techniques.

\paragraph{\textit{Poster Design}}
The poster design task involved three application windows (Figure~\ref{fig:study_task}b). The center window served as the main canvas where participants created a poster, with visual guides such as text-box outlines, image placeholders, and a designated signature area. The left and right windows provided text and image materials for the design. Participants first searched for appropriate text and images, then dragged them onto the center design canvas. After placing the content, they performed fine-grained adjustments such as translation, rotation, and scaling to precisely position the 2D elements within the required layout. Finally, participants signed their initials and drew additional shapes to personalize their poster. This design scenario was selected to allow participants to extensively explore fine-grained multi-touch interactions as well as precise cursor and stroke input using the HandPad techniques.

\subsection{Results}
To assess the HandPad system, which utilizes asymmetric bimanual interaction and a tangible hand surface with self-haptics and hand remapping, we employed questionnaires to evaluate participants’ perceptions of ownership, agency, and tactile sensation~\cite{gonzalez2018avatar}. Furthermore, we evaluated the usability of the HandPad system using additional customized questions that examined the effectiveness of each interaction technique. Finally, we conducted post-study interviews with each participant to gather qualitative feedback on the HandPad system, during which we also asked them to compare their experience using HandPad with the common mid-air gestures used in current VR headsets.

\subsubsection{Ownership, Agency, and Tactile Sensation}
HandPad remaps both hands to interact with large and distant windows while leveraging the NDH as a tangible interactive surface to support comfortable interactions. However, this remapping may influence participants’ perceptions of ownership and agency over the remapped hands. In addition, it is unclear whether touching the palm of the hand can produce a convincing tactile sensation when interacting with a virtual window. To examine these aspects, we selected relevant questions from the embodiment questionnaire~\cite{gonzalez2018avatar}, including Q1–2 for Ownership, Q6–9 for Agency, and Q10–13 for Tactile Sensation. Responses were collected using a 5-point Likert scale ranging from ``strongly disagree'' to ``strongly agree''.

Participants generally reported strong ownership (Median = 1, IQR = 1) and agency (Median = 2, IQR = 1) over their remapped virtual hands. Participants also reported a strong tactile sensation when interacting with windows using the NDH surface (Median = 2, IQR = 1), suggesting that touching the palm can effectively approximate the feeling of interacting with virtual windows. 
Several participants emphasized the sense of control provided by the technique. For example, P7 noted, ``\textit{I felt more in control of what was happening with the HandPad.}''. Similarly, P9 stated, ``\textit{HandPad gave me a sense of control over the window and the stuff I was selecting.}'' They liked that they could also feel the tactile feedback when performing gestures that use the hand as a touch interface. Participants also appreciated the tactile feedback provided by the hand surface when performing gestures that used the hand as a touch interface. For instance, P4 commented, ``\textit{It actually felt like I was touching something when I tapped on my palm, which made the interaction more natural.}''

\subsubsection{Usability}
\begin{figure}[tb]
 \centering
 \includegraphics[width=\linewidth]{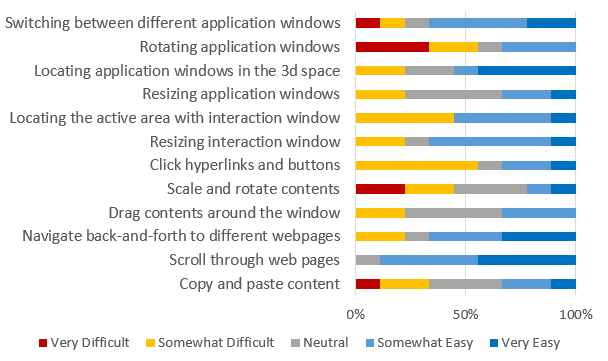}
 \vspace{-6mm}
 \caption{Ease of Use for content, interaction window, and application window interactions. \vspace{-3mm}}
 \label{fig:usability}
\end{figure}

While participants generally found the HandPad interaction techniques to be learnable and practical, their feedback on the usability of individual features varied (Figure~\ref{fig:usability}).

\paragraph{\textit{Interaction Window}}
Manipulating the interaction window was rated largely positively. In the \textbf{Semi-Snap} mode, participants found it easy to resize the interaction window (Median = 4, IQR = 1) and to locate the active interaction area using the interaction window (Median = 4, IQR = 2). Most participants appreciated the flexibility of adjusting the interaction window to accommodate large or distant windows. As P5 noted, ``\textit{The pad size can be adjusted and placed everywhere. Besides, the pad can be fixed or moved according to my demand.}''

\paragraph{\textit{Touch Interactions on a Window}}
In the \textbf{Snap} mode, most participants found it easy to perform interactions on a window, including single-touch interactions such as navigating back and forth between webpages (Median = 4, IQR = 2) and dragging content within a window (Median = 3, IQR = 1). Multi-touch interactions were also rated positively, including scrolling through webpages (Median = 4, IQR = 1) and performing more precise manipulations such as scaling and rotating 2D content (Median = 3, IQR = 1), which ``\textit{require some effort to achieve the correct position and orientation}'' (P7). In the \textbf{Semi-Snap} mode, copying and pasting 2D content by dragging and dropping it across multiple windows was also rated as acceptable (Median = 3, IQR = 2). Surprisingly, some participants reported some difficulty when clicking hyperlinks and buttons (Median = 2, IQR = 2), especially when compared with mid-air selection. Several participants mentioned that using the Semi-Snap mode to locate the interaction area and then switching to Snap mode to perform the selection introduced additional steps. 

\paragraph{\textit{Window Management}}
Regarding manipulation and management of application windows in the \textbf{Over-Snap} mode, most participants found it easy to locate application windows in 3D space (Median = 4, IQR = 2), resize application windows (Median = 3, IQR =1), and switch between different application windows (Median = 4, IQR = 1). However, participants encountered some difficulties when rotating application windows (Median = 2, IQR = 3). As P1 noted, this may be attributed to the high dexterity of hand motion, which introduces ``\textit{too much degree of freedom for [window] rotation}'' (P1) and makes it harder to control precisely. 

\subsubsection{Qualitative Feedback on the Interaction Techniques}

\paragraph{\textit{Bimanual Interactions}}
Most participants (7/9) appreciated that using the NDH to establish a spatial frame and enable different interaction contexts through Snap mode gestures was intuitive, efficient, and easy to learn. They noted that the mode gestures were easy to perform and distinguish from one another. Participants also appreciated the Semi-Snap mode for controlling the interaction window and the Snap mode for stabilized interaction. As P1 noted, ``\textit{It helped me to switch between the active and focused interaction area on the large window.}'' Participants further valued the seamless transition between interaction modes. For example, P7 mentioned, ``\textit{I like that the gesture (Semi-Snap) is a step before the Snap because it is very easy to go from Semi-Snap to Snap and come back.}'' Several participants also enjoyed being able to touch the NDH with the DH ``\textit{just like a touchpad.}''(P4), which supported more precise and sustained touch interactions. Regarding window management, most participants (6/9) appreciated the application window resizing and repositioning features provided by the Over-Snap mode. For instance, P9 noted that ``\textit{it allowed me to reposition the window closer to my eyes.}''. However, dragging application windows across long distances by only moving the NDH could be somewhat challenging and ``\textit{needed a lot of attempts.}'' (P7). In such cases, participants found that using complementary DH mid-air gestures anchored to the NDH was more efficient. 

\paragraph{\textit{Hand Remapping}} Most participants (8/9) appreciated that HandPad allowed them to rest or lower their hands during interaction, indicating the ergonomic benefit of hand remapping. For example, P5 reported, ``\textit{With the Snap mode, I can rest my arm, so it's more comfortable for me to operate.}'' However, participants may require practice or reminders to fully leverage the ability to reposition their hands and reduce fatigue during extended interaction. Some participants mentioned that they would forget to relax their hands when they became highly engaged in the task, only noticing the strain afterward. For instance, P4 mentioned, ``\textit{It mitigates physical exertion if I remember that I can freely use my hand, but I sometimes forgot that.}'' On the other hand, if participants lowered their hands too much, they could fall outside the camera’s tracking range, which might interrupt the interaction flow. This sometimes resulted in repeated attempts and increased fatigue. As P2 explained, ``\textit{Sometimes [tracking] is not accurate, so I need to do the same thing multiple times.}''

\paragraph{\textit{Comparison with Controllers and Mid-Air Gestures}}
When comparing their experience using controllers with HandPad, participants perceived clear benefits of HandPad for 2D window interactions. Several participants noted that the hand remapping made the interaction feel more natural and intuitive than controller-based input. For example, P2 stated, ``\textit{I like how it imitates my hand movement, and it's great to do activities on a big screen without controllers. It somehow feels like I am using a touch screen with feedback…}'' Participants (4/9) who were already familiar with interacting with 2D windows using mid-air gestures particularly appreciated the multi-touch and cross-window interactions supported by HandPad. They noted that these interactions felt more fluid and easy to control compared to traditional mid-air gestures, which often lack a stable physical reference.

\section{DISCUSSION}
HandPad provides a promising approach for supporting productivity-oriented 2D window interactions in VR. Participants appreciated that asymmetric bimanual coordination enabled smooth mode switching and efficient window manipulation. In addition, participants reported strong perceptions of comfort, ownership, and agency over the remapped virtual hands. They also appreciated that interacting with the NDH surface provided convincing tactile feedback, allowing the NDH to function as a tangible interaction surface on which the DH can perform fine-grained touch interactions. 

\subsection{NDH Gestures for Contextual Mode Switching}
HandPad uses NDH gestures to transition between different interaction contexts. The Semi-Snap, Snap, and Over-Snap modes support a layered interaction structure that allows users to progressively move from across-window navigation to precise content manipulation and finally to application-level window management. Participants appreciated the clear separation between these modes and the smooth transitions between them. They also found the gestures convenient and easy to perform while allowing natural hand movement. These findings suggest that assigning the NDH to establish interaction context enables intuitive mode switching while allowing the dominant hand to focus on fine-grained manipulation, which aligns with principles of asymmetric bimanual interaction \cite{guiard1987asymmetric}.

\subsection{NDH as a Tangible Interaction Surface}
HandPad also leverages the NDH as a tangible interaction surface that establishes spatial frame reference, stabilizes input and enables touch-based gestures from the DH. Participants frequently described the interaction as similar to using a touchpad or touchscreen, which enabled more stable and precise control compared to purely mid-air gestures. By providing passive haptic feedback and a spatial reference frame for the DH to perform touch interactions, the NDH surface helps mitigate common issues in mid-air interaction such as hand jitter and lack of tactile feedback. This finding aligns with prior work showing that physical constraints can improve interaction precision by reducing unintended movement~\cite{10.1145/3025453.3025474}. In addition, participants reported that touch interactions on the NDH palm were particularly useful for sustained tasks such as scrolling, dragging content, and performing multi-touch manipulations. Participants also reported that this coordination felt intuitive and easy to learn, suggesting that asymmetric bimanual interaction can potentially support complex window manipulation tasks in VR.

\subsection{Combining Hand-Surface and Mid-Air Input}
While participants generally appreciated touch interactions on the NDH surface, some lightweight actions, such as clicking, were perceived as more efficient with direct mid-air selection. Several participants noted that using Semi-Snap to locate the interaction area and then switching to Snap for selection introduced additional steps for simple actions. In these cases, mid-air gestures provided a faster alternative without requiring interaction window repositioning. This feedback suggests that while hand-surface interactions are effective for precision-demanding or sustained manipulation tasks, lightweight selections may benefit from more direct input methods. Hybrid interaction strategies may therefore better support the diverse requirements of VR productivity workflows. For example, hand-surface interaction is beneficial when precision and control are required or when interactions are sustained, whereas mid-air gestures are better suited for frequent and brief interactions.

\subsection{Ergonomics of Hand Remapping}
HandPad mitigates the “gorilla arm” effect by allowing users to perform interactions at a lower and more comfortable physical position while remapping their hands to higher virtual windows. Many participants appreciated this feature and reported that it allowed them to rest or lower their arms during interaction. 
Our study also revealed that users sometimes forgot to reposition their hands during prolonged tasks, particularly when they were deeply engaged in the activity. These findings suggest that the ergonomic benefits of input remapping may require a period of adaptation, as users need to actively take advantage of the flexibility to reposition their hands for more comfortable interaction.

\subsection{Limitations and Future Work}
While participants appreciated the NDH mode gestures, the flexibility of hand motion sometimes resulted in excessive degrees of freedom during window manipulation. For example, rotating application windows using the NDH in the Over-Snap mode was occasionally perceived as difficult to control. This suggests that such interactions may benefit from additional constraints, such as axis-locking mechanisms or stabilization techniques, to reduce unintended motion during manipulation. In addition, our current gesture recognition uses fixed thresholds for joint flexion and inter-finger angles across all participants. In practice, these gestures could be further improved by adapting hand models to individual users. Personalized calibration could help accommodate variations in hand size and finger stretchability.

Our system employs a lightweight machine learning model for device-free touch detection on the hand surface. Although participants did not report noticeable inaccuracies during the study, future work could explore more advanced vision-based models to further improve touch detection robustness. For example, recent work has leveraged vision transformers for on-skin touch detection~\cite{mollyn2024egotouch}.

While HandPad allows users to lower their hands through input remapping, users need time to adapt to this technique and may forget to do so if they're too focused on the task. To further improve the ergonomics of prolonged bare-hand interaction, future systems can provide subtle visual cues that encourage users to maintain comfortable postures during interaction. In addition, the current limitations of optical hand tracking may restrict how far users can relax their hands before tracking quality degrades. Future systems could address this issue through improved tracking robustness.

\section{CONCLUSION}
We presented HandPad, a set of bare-hand bimanual interaction techniques that support multi-window interaction in VR through asymmetric hand roles, self-haptic feedback, and input remapping. By using the non-dominant hand as a spatial frame and tangible interaction surface, HandPad enables the dominant hand to perform precise manipulations on virtual windows. 
Results from a user study show that participants perceived strong ownership and agency over the remapped hands and found the HandPad interaction techniques intuitive and effective for productivity-oriented tasks. 
Participants particularly appreciated the stability provided by the hand surface and the ability to perform fluid multi-touch and cross-window interactions. Our findings further highlight the potential of combining hand-surface interaction with complementary mid-air gestures to support efficient and comfortable interaction with large virtual workspaces in VR.


\bibliographystyle{unsrt}

\bibliography{reference}

@article{guiard1987asymmetric,
  title={Asymmetric division of labor in human skilled bimanual action: The kinematic chain as a model},
  author={Guiard, Yves},
  journal={Journal of motor behavior},
  volume={19},
  number={4},
  pages={486--517},
  year={1987},
  publisher={Taylor \& Francis}
}

@inproceedings{hinckley1997cooperative,
author = {Hinckley, Ken and Pausch, Randy and Proffitt, Dennis and Patten, James and Kassell, Neal},
title = {Cooperative Bimanual Action},
year = {1997},
isbn = {0897918029},
publisher = {Association for Computing Machinery},
address = {New York, NY, USA},
url = {https://doi.org/10.1145/258549.258571},
doi = {10.1145/258549.258571},
booktitle = {Proceedings of the ACM SIGCHI Conference on Human Factors in Computing Systems},
pages = {27–34},
numpages = {8},
location = {Atlanta, Georgia, USA},
series = {CHI '97}
}

@inproceedings{hinckley1997attention,
author = {Hinckley, Ken and Pausch, Randy and Proffitt, Dennis},
title = {Attention and visual feedback: the bimanual frame of reference},
year = {1997},
isbn = {0897918843},
publisher = {Association for Computing Machinery},
address = {New York, NY, USA},
url = {https://doi.org/10.1145/253284.253318},
doi = {10.1145/253284.253318},
booktitle = {Proceedings of the 1997 Symposium on Interactive 3D Graphics},
pages = {121–ff.},
location = {Providence, Rhode Island, USA},
series = {I3D '97}
}

@inproceedings{webb2016wearables,
author = {Webb, Andrew M. and Pahud, Michel and Hinckley, Ken and Buxton, Bill},
title = {Wearables as Context for Guiard-abiding Bimanual Touch},
year = {2016},
isbn = {9781450341899},
publisher = {Association for Computing Machinery},
address = {New York, NY, USA},
url = {https://doi.org/10.1145/2984511.2984564},
doi = {10.1145/2984511.2984564},
booktitle = {Proceedings of the 29th Annual Symposium on User Interface Software and Technology},
pages = {287–300},
numpages = {14},
location = {Tokyo, Japan},
series = {UIST '16}
}

@inproceedings{kabbash1994two,
author = {Kabbash, Paul and Buxton, William and Sellen, Abigail},
title = {Two-Handed Input in a Compound Task},
year = {1994},
isbn = {0897916506},
publisher = {Association for Computing Machinery},
address = {New York, NY, USA},
url = {https://doi.org/10.1145/191666.191808},
doi = {10.1145/191666.191808},
booktitle = {Proceedings of the SIGCHI Conference on Human Factors in Computing Systems},
pages = {417–423},
numpages = {7},
location = {Boston, Massachusetts, USA},
series = {CHI '94}
}

@inproceedings{role1999balakrishnan,
author = {Balakrishnan, Ravin and Hinckley, Ken},
title = {The Role of Kinesthetic Reference Frames in Two-Handed Input Performance},
year = {1999},
isbn = {1581130759},
publisher = {Association for Computing Machinery},
address = {New York, NY, USA},
url = {https://doi.org/10.1145/320719.322599},
doi = {10.1145/320719.322599},
booktitle = {Proceedings of the 12th Annual ACM Symposium on User Interface Software and Technology},
pages = {171–178},
numpages = {8},
location = {Asheville, North Carolina, USA},
series = {UIST '99}
}

@inproceedings{hinckley1994passive,
author = {Hinckley, Ken and Pausch, Randy and Goble, John C. and Kassell, Neal F.},
title = {Passive Real-World Interface Props for Neurosurgical Visualization},
year = {1994},
isbn = {0897916506},
publisher = {Association for Computing Machinery},
address = {New York, NY, USA},
url = {https://doi.org/10.1145/191666.191821},
doi = {10.1145/191666.191821},
booktitle = {Proceedings of the SIGCHI Conference on Human Factors in Computing Systems},
pages = {452–458},
numpages = {7},
location = {Boston, Massachusetts, USA},
series = {CHI '94}
}

@inproceedings{two1994kabbash,
author = {Kabbash, Paul and Buxton, William and Sellen, Abigail},
title = {Two-Handed Input in a Compound Task},
year = {1994},
isbn = {0897916506},
publisher = {Association for Computing Machinery},
address = {New York, NY, USA},
url = {https://doi.org/10.1145/191666.191808},
doi = {10.1145/191666.191808},
booktitle = {Proceedings of the SIGCHI Conference on Human Factors in Computing Systems},
pages = {417–423},
numpages = {7},
location = {Boston, Massachusetts, USA},
series = {CHI '94}
}

@InProceedings{hansberger2017dispelling,
author="Hansberger, Jeffrey T.
and Peng, Chao
and Mathis, Shannon L.
and Areyur Shanthakumar, Vaidyanath
and Meacham, Sarah C.
and Cao, Lizhou
and Blakely, Victoria R.",
editor="Lackey, Stephanie
and Chen, Jessie",
title="Dispelling the Gorilla Arm Syndrome: The Viability of Prolonged Gesture Interactions",
booktitle="Virtual, Augmented and Mixed Reality",
year="2017",
publisher="Springer International Publishing",
address="Cham",
pages="505--520",
isbn="978-3-319-57987-0"
}

@inproceedings{medeiros2023benefits,
author = {Medeiros, Daniel and Wilson, Graham and Mcgill, Mark and Brewster, Stephen Anthony},
title = {The Benefits of Passive Haptics and Perceptual Manipulation for Extended Reality Interactions in Constrained Passenger Spaces},
year = {2023},
isbn = {9781450394215},
publisher = {Association for Computing Machinery},
address = {New York, NY, USA},
url = {https://doi.org/10.1145/3544548.3581079},
doi = {10.1145/3544548.3581079},
booktitle = {Proceedings of the 2023 CHI Conference on Human Factors in Computing Systems},
articleno = {232},
numpages = {19},
location = {Hamburg, Germany},
series = {CHI '23}
}

@inproceedings{feuchtner2018ownershift,
author = {Feuchtner, Tiare and M\"{u}ller, J\"{o}rg},
title = {Ownershift: Facilitating Overhead Interaction in Virtual Reality with an Ownership-Preserving Hand Space Shift},
year = {2018},
isbn = {9781450359481},
publisher = {Association for Computing Machinery},
address = {New York, NY, USA},
url = {https://doi.org/10.1145/3242587.3242594},
doi = {10.1145/3242587.3242594},
booktitle = {Proceedings of the 31st Annual ACM Symposium on User Interface Software and Technology},
pages = {31–43},
numpages = {13},
location = {Berlin, Germany},
series = {UIST '18}
}

@inproceedings{lindeman1999towards,
author = {Lindeman, Robert W. and Sibert, John L. and Hahn, James K.},
title = {Towards usable VR: an empirical study of user interfaces for immersive virtual environments},
year = {1999},
isbn = {0201485591},
publisher = {Association for Computing Machinery},
address = {New York, NY, USA},
url = {https://doi.org/10.1145/302979.302995},
doi = {10.1145/302979.302995},
booktitle = {Proceedings of the SIGCHI Conference on Human Factors in Computing Systems},
pages = {64–71},
numpages = {8},
location = {Pittsburgh, Pennsylvania, USA},
series = {CHI '99}
}

@inproceedings{casiez20121,
author = {Casiez, G\'{e}ry and Roussel, Nicolas and Vogel, Daniel},
title = {1 € filter: a simple speed-based low-pass filter for noisy input in interactive systems},
year = {2012},
isbn = {9781450310154},
publisher = {Association for Computing Machinery},
address = {New York, NY, USA},
url = {https://doi.org/10.1145/2207676.2208639},
doi = {10.1145/2207676.2208639},
booktitle = {Proceedings of the SIGCHI Conference on Human Factors in Computing Systems},
pages = {2527–2530},
numpages = {4},
location = {Austin, Texas, USA},
series = {CHI '12}
}

@inproceedings{10.1145/1622176.1622205,
author = {McCallum, David C. and Irani, Pourang},
title = {ARC-Pad: Absolute+relative Cursor Positioning for Large Displays with a Mobile Touchscreen},
year = {2009},
isbn = {9781605587455},
publisher = {Association for Computing Machinery},
address = {New York, NY, USA},
url = {https://doi.org/10.1145/1622176.1622205},
doi = {10.1145/1622176.1622205},
booktitle = {Proceedings of the 22nd Annual ACM Symposium on User Interface Software and Technology},
pages = {153–156},
numpages = {4},
location = {Victoria, BC, Canada},
series = {UIST '09}
}

@inproceedings{10.1145/2470654.2470773,
author = {Nancel, Mathieu and Chapuis, Olivier and Pietriga, Emmanuel and Yang, Xing-Dong and Irani, Pourang P. and Beaudouin-Lafon, Michel},
title = {High-Precision Pointing on Large Wall Displays Using Small Handheld Devices},
year = {2013},
isbn = {9781450318990},
publisher = {Association for Computing Machinery},
address = {New York, NY, USA},
url = {https://doi.org/10.1145/2470654.2470773},
doi = {10.1145/2470654.2470773},
booktitle = {Proceedings of the SIGCHI Conference on Human Factors in Computing Systems},
pages = {831–840},
numpages = {10},
location = {Paris, France},
series = {CHI '13}
}

@inproceedings{10.1145/3173574.3173919,
author = {Knierim, Pascal and Schwind, Valentin and Feit, Anna Maria and Nieuwenhuizen, Florian and Henze, Niels},
title = {Physical Keyboards in Virtual Reality: Analysis of Typing Performance and Effects of Avatar Hands},
year = {2018},
isbn = {9781450356206},
publisher = {Association for Computing Machinery},
address = {New York, NY, USA},
url = {https://doi.org/10.1145/3173574.3173919},
doi = {10.1145/3173574.3173919},
booktitle = {Proceedings of the 2018 CHI Conference on Human Factors in Computing Systems},
pages = {1–9},
numpages = {9},
location = {Montreal QC, Canada},
series = {CHI '18}
}

@ARTICLE{8617763,
  author={Grubert, Jens and Ofek, Eyal and Pahud, Michel and Kristensson, Per Ola},
  journal={IEEE Computer Graphics and Applications}, 
  title={The Office of the Future: Virtual, Portable, and Global}, 
  year={2018},
  volume={38},
  number={6},
  pages={125-133},
  doi={10.1109/MCG.2018.2875609}}

@inproceedings{10.1145/3313831.3376724,
author = {Ruvimova, Anastasia and Kim, Junhyeok and Fritz, Thomas and Hancock, Mark and Shepherd, David C.},
title = {"Transport Me Away": Fostering Flow in Open Offices through Virtual Reality},
year = {2020},
isbn = {9781450367080},
publisher = {Association for Computing Machinery},
address = {New York, NY, USA},
url = {https://doi.org/10.1145/3313831.3376724},
doi = {10.1145/3313831.3376724},
booktitle = {Proceedings of the 2020 CHI Conference on Human Factors in Computing Systems},
pages = {1–14},
numpages = {14},
location = {Honolulu, HI, USA},
series = {CHI '20}
}

@inproceedings{10.1145/3290605.3300243,
author = {Surale, Hemant Bhaskar and Gupta, Aakar and Hancock, Mark and Vogel, Daniel},
title = {TabletInVR: Exploring the Design Space for Using a Multi-Touch Tablet in Virtual Reality},
year = {2019},
isbn = {9781450359702},
publisher = {Association for Computing Machinery},
address = {New York, NY, USA},
url = {https://doi.org/10.1145/3290605.3300243},
doi = {10.1145/3290605.3300243},
booktitle = {Proceedings of the 2019 CHI Conference on Human Factors in Computing Systems},
pages = {1–13},
numpages = {13},
location = {Glasgow, Scotland Uk},
series = {CHI '19}
}

@inproceedings{10.1145/3313831.3376652,
author = {Jetter, Hans-Christian and R\"{a}dle, Roman and Feuchtner, Tiare and Anthes, Christoph and Friedl, Judith and Klokmose, Clemens Nylandsted},
title = {"In VR, Everything is Possible!": Sketching and Simulating Spatially-Aware Interactive Spaces in Virtual Reality},
year = {2020},
isbn = {9781450367080},
publisher = {Association for Computing Machinery},
address = {New York, NY, USA},
url = {https://doi.org/10.1145/3313831.3376652},
doi = {10.1145/3313831.3376652},
booktitle = {Proceedings of the 2020 CHI Conference on Human Factors in Computing Systems},
pages = {1–16},
numpages = {16},
location = {Honolulu, HI, USA},
series = {CHI '20}
}

@INPROCEEDINGS{9995404,
  author={Cheng, Yi Fei and Luong, Tiffany and Fender, Andreas Rene and Streli, Paul and Holz, Christian},
  booktitle={2022 IEEE International Symposium on Mixed and Augmented Reality (ISMAR)}, 
  title={ComforTable User Interfaces: Surfaces Reduce Input Error, Time, and Exertion for Tabletop and Mid-air User Interfaces}, 
  year={2022},
  volume={},
  number={},
  pages={150-159},
  doi={10.1109/ISMAR55827.2022.00029}}

@misc{ofek2020practical,
      title={Towards a Practical Virtual Office for Mobile Knowledge Workers}, 
      author={Eyal Ofek and Jens Grubert and Michel Pahud and Mark Phillips and Per Ola Kristensson},
      year={2020},
      eprint={2009.02947},
      archivePrefix={arXiv},
      primaryClass={cs.HC}
}

@ARTICLE{9212653,
  author={Biener, Verena and Schneider, Daniel and Gesslein, Travis and Otte, Alexander and Kuth, Bastian and Kristensson, Per Ola and Ofek, Eyal and Pahud, Michel and Grubert, Jens},
  journal={IEEE Transactions on Visualization and Computer Graphics}, 
  title={Breaking the Screen: Interaction Across Touchscreen Boundaries in Virtual Reality for Mobile Knowledge Workers}, 
  year={2020},
  volume={26},
  number={12},
  pages={3490-3502},
  doi={10.1109/TVCG.2020.3023567}}

@INPROCEEDINGS{9284682,
  author={Gesslein, Travis and Biener, Verena and Gagel, Philipp and Schneider, Daniel and Kristensson, Per Ola and Ofek, Eyal and Pahud, Michel and Grubert, Jens},
  booktitle={2020 IEEE International Symposium on Mixed and Augmented Reality (ISMAR)}, 
  title={Pen-based Interaction with Spreadsheets in Mobile Virtual Reality}, 
  year={2020},
  volume={},
  number={},
  pages={361-373},
  doi={10.1109/ISMAR50242.2020.00063}}

@inproceedings{10.1145/3025453.3025474,
author = {Arora, Rahul and Kazi, Rubaiat Habib and Anderson, Fraser and Grossman, Tovi and Singh, Karan and Fitzmaurice, George},
title = {Experimental Evaluation of Sketching on Surfaces in VR},
year = {2017},
isbn = {9781450346559},
publisher = {Association for Computing Machinery},
address = {New York, NY, USA},
url = {https://doi.org/10.1145/3025453.3025474},
doi = {10.1145/3025453.3025474},
booktitle = {Proceedings of the 2017 CHI Conference on Human Factors in Computing Systems},
pages = {5643–5654},
numpages = {12},
location = {Denver, Colorado, USA},
series = {CHI ’17}
}

@inproceedings{10.1145/3267782.3267788,
author = {Wacker, Philipp and Wagner, Adrian and Voelker, Simon and Borchers, Jan},
title = {Physical Guides: An Analysis of 3D Sketching Performance on Physical Objects in Augmented Reality},
year = {2018},
isbn = {9781450357081},
publisher = {Association for Computing Machinery},
address = {New York, NY, USA},
url = {https://doi.org/10.1145/3267782.3267788},
doi = {10.1145/3267782.3267788},
booktitle = {Proceedings of the Symposium on Spatial User Interaction},
pages = {25–35},
numpages = {11},
location = {Berlin, Germany},
series = {SUI '18}
}

@article{fujita2023human,
  title={Human-Workspace Interaction: prior research efforts and future challenges for supporting knowledge workers},
  author={Fujita, Kazuyuki and Takashima, Kazuki and Itoh, Yuichi and Kitamura, Yoshifumi},
  journal={Quality and User Experience},
  volume={8},
  number={1},
  pages={7},
  year={2023},
  publisher={Springer}
}

@book{simeone2023everyday,
  title={Everyday Virtual and Augmented Reality},
  author={Simeone, Adalberto and Weyers, Benjamin and Bialkova, Svetlana and Lindeman, Robert W},
  year={2023},
  publisher={Springer Nature}
}

@INPROCEEDINGS{8943608,
  author={Guo, Jie and Weng, Dongdong and Zhang, Zhenliang and Jiang, Haiyan and Liu, Yue and Wang, Yongtian and Duh, Henry Been-Lirn},
  booktitle={2019 IEEE International Symposium on Mixed and Augmented Reality (ISMAR)}, 
  title={Mixed Reality Office System Based on Maslow’s Hierarchy of Needs: Towards the Long-Term Immersion in Virtual Environments}, 
  year={2019},
  volume={},
  number={},
  pages={224-235},
  doi={10.1109/ISMAR.2019.00019}}

@inproceedings{ying2026redirected,
author = {Ying, Wen and Kim, Yeonsu and Rahman, Adil and Hu, Erzhen and Lee, Geehyuk and Heo, Seongkook},
title = {Redirected Pinch: Efficient and Comfortable Bare-Hand Interaction for 2D Windows in VR},
year = {2026},
isbn = {9798400722783},
publisher = {Association for Computing Machinery},
address = {New York, NY, USA},
url = {https://doi.org/10.1145/3772318.3791512},
doi = {10.1145/3772318.3791512},
booktitle = {Proceedings of the 2026 CHI Conference on Human Factors in Computing Systems},
articleno = {1686},
numpages = {18},
location = {
},
series = {CHI '26}
}

@INPROCEEDINGS{4142852,
  author={Andujar, Carlos and Argelaguet, Ferran},
  booktitle={2007 IEEE Symposium on 3D User Interfaces}, 
  title={Virtual Pads: Decoupling Motor Space and Visual Space for Flexible Manipulation of 2D Windows within VEs}, 
  year={2007},
  volume={},
  number={},
  pages={},
  doi={10.1109/3DUI.2007.340781}}

@INPROCEEDINGS{9284722,
  author={Brasier, Eugenie and Chapuis, Olivier and Ferey, Nicolas and Vezien, Jeanne and Appert, Caroline},
  booktitle={2020 IEEE International Symposium on Mixed and Augmented Reality (ISMAR)}, 
  title={ARPads: Mid-air Indirect Input for Augmented Reality}, 
  year={2020},
  volume={},
  number={},
  pages={332-343},
  doi={10.1109/ISMAR50242.2020.00060}}

@inproceedings{wentzel2020improving,
author = {Wentzel, Johann and d'Eon, Greg and Vogel, Daniel},
title = {Improving Virtual Reality Ergonomics Through Reach-Bounded Non-Linear Input Amplification},
year = {2020},
isbn = {9781450367080},
publisher = {Association for Computing Machinery},
address = {New York, NY, USA},
url = {https://doi.org/10.1145/3313831.3376687},
doi = {10.1145/3313831.3376687},
booktitle = {Proceedings of the 2020 CHI Conference on Human Factors in Computing Systems},
pages = {1–12},
numpages = {12},
location = {Honolulu, HI, USA},
series = {CHI '20}
}

@ARTICLE{iqbal2021reducing,
  author={Iqbal, Hasan and Latif, Seemab and Yan, Yukang and Yu, Chun and Shi, Yuanchun},
  journal={IEEE Access}, 
  title={Reducing Arm Fatigue in Virtual Reality by Introducing 3D-Spatial Offset}, 
  year={2021},
  volume={9},
  number={},
  pages={64085-64104},
  doi={10.1109/ACCESS.2021.3075769}}

@inproceedings{hirzle2020survey,
author = {Hirzle, Teresa and Cordts, Maurice and Rukzio, Enrico and Bulling, Andreas},
title = {A Survey of Digital Eye Strain in Gaze-Based Interactive Systems},
year = {2020},
isbn = {9781450371339},
publisher = {Association for Computing Machinery},
address = {New York, NY, USA},
url = {https://doi.org/10.1145/3379155.3391313},
doi = {10.1145/3379155.3391313},
booktitle = {ACM Symposium on Eye Tracking Research and Applications},
articleno = {9},
numpages = {12},
location = {Stuttgart, Germany},
series = {ETRA '20 Full Papers}
}

@inproceedings{10.1145/3461778.3462076,
author = {Le, Khanh-Duy and Tran, Tanh Quang and Chlasta, Karol and Krejtz, Krzysztof and Fjeld, Morten and Kunz, Andreas},
title = {VXSlate: Exploring Combination of Head Movements and Mobile Touch for Large Virtual Display Interaction},
year = {2021},
isbn = {9781450384766},
publisher = {Association for Computing Machinery},
address = {New York, NY, USA},
url = {https://doi.org/10.1145/3461778.3462076},
doi = {10.1145/3461778.3462076},
booktitle = {Proceedings of the 2021 ACM Designing Interactive Systems Conference},
pages = {283–297},
numpages = {15},
location = {Virtual Event, USA},
series = {DIS '21}
}

@book{jones2006human,
    author = {Jones, Lynette A. and Lederman, Susan J.},
    title = "{Human Hand Function}",
    publisher = {Oxford University Press},
    year = {2006},
    month = {05},
    isbn = {9780195173154},
    doi = {10.1093/acprof:oso/9780195173154.001.0001},
    url = {https://doi.org/10.1093/acprof:oso/9780195173154.001.0001},
}

@inproceedings{10.1145/3472749.3474810,
author = {Fang, Cathy Mengying and Harrison, Chris},
title = {Retargeted Self-Haptics for Increased Immersion in VR without Instrumentation},
year = {2021},
isbn = {9781450386357},
publisher = {Association for Computing Machinery},
address = {New York, NY, USA},
url = {https://doi.org/10.1145/3472749.3474810},
doi = {10.1145/3472749.3474810},
booktitle = {The 34th Annual ACM Symposium on User Interface Software and Technology},
pages = {1109–1121},
numpages = {13},
location = {Virtual Event, USA},
series = {UIST '21}
}

@inproceedings{mollyn2024egotouch,
author = {Mollyn, Vimal and Harrison, Chris},
title = {EgoTouch: On-Body Touch Input Using AR/VR Headset Cameras},
year = {2024},
isbn = {9798400706288},
publisher = {Association for Computing Machinery},
address = {New York, NY, USA},
url = {https://doi.org/10.1145/3654777.3676455},
doi = {10.1145/3654777.3676455},
booktitle = {Proceedings of the 37th Annual ACM Symposium on User Interface Software and Technology},
articleno = {69},
numpages = {11},
location = {Pittsburgh, PA, USA},
series = {UIST '24}
}

@inproceedings{10.1145/3313831.3376132,
author = {Wang, Bryan and Grossman, Tovi},
title = {BlyncSync: Enabling Multimodal Smartwatch Gestures with Synchronous Touch and Blink},
year = {2020},
isbn = {9781450367080},
publisher = {Association for Computing Machinery},
address = {New York, NY, USA},
url = {https://doi.org/10.1145/3313831.3376132},
doi = {10.1145/3313831.3376132},
booktitle = {Proceedings of the 2020 CHI Conference on Human Factors in Computing Systems},
pages = {1–14},
numpages = {14},
location = {Honolulu, HI, USA},
series = {CHI '20}
}

@inproceedings{10.1145/3491102.3501898,
author = {Pei, Siyou and Chen, Alexander and Lee, Jaewook and Zhang, Yang},
title = {Hand Interfaces: Using Hands to Imitate Objects in AR/VR for Expressive Interactions},
year = {2022},
isbn = {9781450391573},
publisher = {Association for Computing Machinery},
address = {New York, NY, USA},
url = {https://doi.org/10.1145/3491102.3501898},
doi = {10.1145/3491102.3501898},
booktitle = {Proceedings of the 2022 CHI Conference on Human Factors in Computing Systems},
articleno = {429},
numpages = {16},
location = {New Orleans, LA, USA},
series = {CHI '22}
}

@inproceedings{surale2019experimental,
author = {Surale, Hemant Bhaskar and Matulic, Fabrice and Vogel, Daniel},
title = {Experimental Analysis of Barehand Mid-air Mode-Switching Techniques in Virtual Reality},
year = {2019},
isbn = {9781450359702},
publisher = {Association for Computing Machinery},
address = {New York, NY, USA},
url = {https://doi.org/10.1145/3290605.3300426},
doi = {10.1145/3290605.3300426},
booktitle = {Proceedings of the 2019 CHI Conference on Human Factors in Computing Systems},
pages = {1–14},
numpages = {14},
location = {Glasgow, Scotland Uk},
series = {CHI '19}
}

@article{hinckley1998two,
  title={Two-handed virtual manipulation},
  author={Hinckley, Ken and Pausch, Randy and Proffitt, Dennis and Kassell, Neal F},
  journal={ACM Transactions on Computer-Human Interaction (TOCHI)},
  volume={5},
  number={3},
  pages={260--302},
  year={1998},
  publisher={ACM New York, NY, USA}
}

@inproceedings{10.1145/258549.258571,
author = {Hinckley, Ken and Pausch, Randy and Proffitt, Dennis and Patten, James and Kassell, Neal},
title = {Cooperative Bimanual Action},
year = {1997},
isbn = {0897918029},
publisher = {Association for Computing Machinery},
address = {New York, NY, USA},
url = {https://doi.org/10.1145/258549.258571},
doi = {10.1145/258549.258571},
booktitle = {Proceedings of the ACM SIGCHI Conference on Human Factors in Computing Systems},
pages = {27–34},
numpages = {8},
location = {Atlanta, Georgia, USA},
series = {CHI '97}
}

@inproceedings{10.5555/1992917.1992939,
author = {Xiao, Robert and Nacenta, Miguel A. and Mandryk, Regan L. and Cockburn, Andy and Gutwin, Carl},
title = {Ubiquitous Cursor: A Comparison of Direct and Indirect Pointing Feedback in Multi-Display Environments},
year = {2011},
isbn = {9781450306935},
publisher = {Canadian Human-Computer Communications Society},
address = {Waterloo, CAN},
booktitle = {Proceedings of Graphics Interface 2011},
pages = {135–142},
numpages = {8},
location = {St. John's, Newfoundland, Canada},
series = {GI '11}
}

@inproceedings{10.1145/1868914.1869036,
author = {Waldner, Manuela and Kruijff, Ernst and Schmalstieg, Dieter},
title = {Bridging Gaps with Pointer Warping in Multi-Display Environments},
year = {2010},
isbn = {9781605589343},
publisher = {Association for Computing Machinery},
address = {New York, NY, USA},
url = {https://doi.org/10.1145/1868914.1869036},
doi = {10.1145/1868914.1869036},
booktitle = {Proceedings of the 6th Nordic Conference on Human-Computer Interaction: Extending Boundaries},
pages = {813–816},
numpages = {4},
location = {Reykjavik, Iceland},
series = {NordiCHI '10}
}

@inproceedings{10.1145/3563657.3596117,
author = {Perella-Holfeld, Francisco and Faleel, Shariff AM and Irani, Pourang},
title = {Evaluating Design Guidelines for Hand Proximate User Interfaces},
year = {2023},
isbn = {9781450398930},
publisher = {Association for Computing Machinery},
address = {New York, NY, USA},
url = {https://doi.org/10.1145/3563657.3596117},
doi = {10.1145/3563657.3596117},
booktitle = {Proceedings of the 2023 ACM Designing Interactive Systems Conference},
pages = {1159–1173},
numpages = {15},
location = {Pittsburgh, PA, USA},
series = {DIS '23}
}

@inproceedings{10.1145/2325616.2325623,
author = {Dezfuli, Niloofar and Khalilbeigi, Mohammadreza and Huber, Jochen and M\"{u}ller, Florian and M\"{u}hlh\"{a}user, Max},
title = {PalmRC: Imaginary Palm-Based Remote Control for Eyes-Free Television Interaction},
year = {2012},
isbn = {9781450311076},
publisher = {Association for Computing Machinery},
address = {New York, NY, USA},
url = {https://doi.org/10.1145/2325616.2325623},
doi = {10.1145/2325616.2325623},
booktitle = {Proceedings of the 10th European Conference on Interactive TV and Video},
pages = {27–34},
numpages = {8},
location = {Berlin, Germany},
series = {EuroITV '12}
}

@inproceedings{10.1145/3332165.3347942,
author = {Arora, Rahul and Kazi, Rubaiat Habib and Kaufman, Danny M. and Li, Wilmot and Singh, Karan},
title = {MagicalHands: Mid-Air Hand Gestures for Animating in VR},
year = {2019},
isbn = {9781450368162},
publisher = {Association for Computing Machinery},
address = {New York, NY, USA},
url = {https://doi.org/10.1145/3332165.3347942},
doi = {10.1145/3332165.3347942},
booktitle = {Proceedings of the 32nd Annual ACM Symposium on User Interface Software and Technology},
pages = {463–477},
numpages = {15},
location = {New Orleans, LA, USA},
series = {UIST '19}
}

@article{gonzalez2018avatar,
  title={Avatar embodiment. towards a standardized questionnaire},
  author={Gonzalez-Franco, Mar and Peck, Tabitha C},
  journal={Frontiers in Robotics and AI},
  volume={5},
  pages={74},
  year={2018},
  publisher={Frontiers Media SA}
}

@inproceedings{10.1145/3313831.3376233,
author = {Zhu, Fengyuan and Grossman, Tovi},
title = {BISHARE: Exploring Bidirectional Interactions Between Smartphones and Head-Mounted Augmented Reality},
year = {2020},
isbn = {9781450367080},
publisher = {Association for Computing Machinery},
address = {New York, NY, USA},
url = {https://doi.org/10.1145/3313831.3376233},
doi = {10.1145/3313831.3376233},
booktitle = {Proceedings of the 2020 CHI Conference on Human Factors in Computing Systems},
pages = {1–14},
numpages = {14},
location = {Honolulu, HI, USA},
series = {CHI '20}
}

@inproceedings{10.1145/1866029.1866033,
author = {Gustafson, Sean and Bierwirth, Daniel and Baudisch, Patrick},
title = {Imaginary Interfaces: Spatial Interaction with Empty Hands and without Visual Feedback},
year = {2010},
isbn = {9781450302715},
publisher = {Association for Computing Machinery},
address = {New York, NY, USA},
url = {https://doi.org/10.1145/1866029.1866033},
doi = {10.1145/1866029.1866033},
booktitle = {Proceedings of the 23nd Annual ACM Symposium on User Interface Software and Technology},
pages = {3–12},
numpages = {10},
location = {New York, New York, USA},
series = {UIST '10}
}

@inproceedings{10.1145/2371574.2371599,
author = {Chen, Xiang 'Anthony' and Marquardt, Nicolai and Tang, Anthony and Boring, Sebastian and Greenberg, Saul},
title = {Extending a Mobile Device's Interaction Space through Body-Centric Interaction},
year = {2012},
isbn = {9781450311052},
publisher = {Association for Computing Machinery},
address = {New York, NY, USA},
url = {https://doi.org/10.1145/2371574.2371599},
doi = {10.1145/2371574.2371599},
booktitle = {Proceedings of the 14th International Conference on Human-Computer Interaction with Mobile Devices and Services},
pages = {151–160},
numpages = {10},
location = {San Francisco, California, USA},
series = {MobileHCI '12}
}

@inproceedings{10.1145/3332165.3347916,
author = {Hayatpur, Devamardeep and Heo, Seongkook and Xia, Haijun and Stuerzlinger, Wolfgang and Wigdor, Daniel},
title = {Plane, Ray, and Point: Enabling Precise Spatial Manipulations with Shape Constraints},
year = {2019},
isbn = {9781450368162},
publisher = {Association for Computing Machinery},
address = {New York, NY, USA},
url = {https://doi.org/10.1145/3332165.3347916},
doi = {10.1145/3332165.3347916},
booktitle = {Proceedings of the 32nd Annual ACM Symposium on User Interface Software and Technology},
pages = {1185–1195},
numpages = {11},
location = {New Orleans, LA, USA},
series = {UIST '19}
}

@inproceedings{10.1145/2858036.2858452,
author = {Kim, Sunjun and Lee, Geehyuk},
title = {TapBoard 2: Simple and Effective Touchpad-like Interaction on a Multi-Touch Surface Keyboard},
year = {2016},
isbn = {9781450333627},
publisher = {Association for Computing Machinery},
address = {New York, NY, USA},
url = {https://doi.org/10.1145/2858036.2858452},
doi = {10.1145/2858036.2858452},
booktitle = {Proceedings of the 2016 CHI Conference on Human Factors in Computing Systems},
pages = {5163–5168},
numpages = {6},
location = {San Jose, California, USA},
series = {CHI '16}
}

@inproceedings{pfeuffer2015gaze,
  title={Gaze-shifting: Direct-indirect input with pen and touch modulated by gaze},
  author={Pfeuffer, Ken and Alexander, Jason and Chong, Ming Ki and Zhang, Yanxia and Gellersen, Hans},
  booktitle={Proceedings of the 28th Annual ACM Symposium on User Interface Software \& Technology},
  pages={373--383},
  year={2015}
}

@inproceedings{10.1145/1385569.1385595,
author = {Brandl, Peter and Forlines, Clifton and Wigdor, Daniel and Haller, Michael and Shen, Chia},
title = {Combining and Measuring the Benefits of Bimanual Pen and Direct-Touch Interaction on Horizontal Interfaces},
year = {2008},
isbn = {9781605581415},
publisher = {Association for Computing Machinery},
address = {New York, NY, USA},
url = {https://doi.org/10.1145/1385569.1385595},
doi = {10.1145/1385569.1385595},
booktitle = {Proceedings of the Working Conference on Advanced Visual Interfaces},
pages = {154–161},
numpages = {8},
location = {Napoli, Italy},
series = {AVI '08}
}

@inproceedings{10.1145/2047196.2047255,
author = {Harrison, Chris and Benko, Hrvoje and Wilson, Andrew D.},
title = {OmniTouch: Wearable Multitouch Interaction Everywhere},
year = {2011},
isbn = {9781450307161},
publisher = {Association for Computing Machinery},
address = {New York, NY, USA},
url = {https://doi.org/10.1145/2047196.2047255},
doi = {10.1145/2047196.2047255},
booktitle = {Proceedings of the 24th Annual ACM Symposium on User Interface Software and Technology},
pages = {441–450},
numpages = {10},
location = {Santa Barbara, California, USA},
series = {UIST '11}
}

@inproceedings{Zhang2019ActiTouch,
author = {Zhang, Yang and Kienzle, Wolf and Ma, Yanjun and Ng, Shiu S. and Benko, Hrvoje and Harrison, Chris},
title = {ActiTouch: Robust Touch Detection for On-Skin AR/VR Interfaces},
year = {2019},
isbn = {9781450368162},
publisher = {Association for Computing Machinery},
address = {New York, NY, USA},
url = {https://doi.org/10.1145/3332165.3347869},
doi = {10.1145/3332165.3347869},
booktitle = {Proceedings of the 32nd Annual ACM Symposium on User Interface Software and Technology},
pages = {1151–1159},
numpages = {9},
location = {New Orleans, LA, USA},
series = {UIST '19}
}

@INPROCEEDINGS{913781,
  author={Bowman, D.A. and Wingrave, C.A.},
  booktitle={Proceedings IEEE Virtual Reality 2001}, 
  title={Design and evaluation of menu systems for immersive virtual environments}, 
  year={2001},
  volume={},
  number={},
  pages={149-156},
  doi={10.1109/VR.2001.913781}}
\end{document}